\documentclass[12pt,reqno]{amsart}

\catcode`\@=11

\thinlines
\newskip\Einheit \Einheit=0.6cm
\newcount\xcoord \newcount\ycoord
\newdimen\xdim \newdimen\ydim \newdimen\PfadD@cke \newdimen\Pfadd@cke
\def\PfadDicke#1{\PfadD@cke#1 \divide\PfadD@cke by2 \Pfadd@cke\PfadD@cke \multiply\PfadD@cke by2}
\long\def\LOOP#1\REPEAT{\def\BODY{#1}\ITERATE}
\def\ITERATE{\BODY \let\next\ITERATE \else\let\next\relax\fi \next}
\let\REPEAT=\fi
\def\Punkt{\hbox{\raise-2pt\hbox to0pt{\hss\scriptsize$\bullet$\hss}}}
\def\DuennPunkt(#1,#2){\unskip
  \raise#2 \Einheit\hbox to0pt{\hskip#1 \Einheit
          \raise-2.5pt\hbox to0pt{\hss\normalsize$\bullet$\hss}\hss}}
\def\NormalPunkt(#1,#2){\unskip
  \raise#2 \Einheit\hbox to0pt{\hskip#1 \Einheit
          \raise-3pt\hbox to0pt{\hss\large$\bullet$\hss}\hss}}
\def\DickPunkt(#1,#2){\unskip
  \raise#2 \Einheit\hbox to0pt{\hskip#1 \Einheit
          \raise-4pt\hbox to0pt{\hss\Large$\bullet$\hss}\hss}}
\def\Kreis(#1,#2){\unskip
  \raise#2 \Einheit\hbox to0pt{\hskip#1 \Einheit
          \raise-4pt\hbox to0pt{\hss\Large$\circ$\hss}\hss}}
\def\Diagonale(#1,#2)#3{\unskip\leavevmode
  \xcoord#1\relax \ycoord#2\relax
      \raise\ycoord \Einheit\hbox to0pt{\hskip\xcoord \Einheit
         \unitlength\Einheit
         \line(1,1){#3}\hss}}
\def\AntiDiagonale(#1,#2)#3{\unskip\leavevmode
  \xcoord#1\relax \ycoord#2\relax \advance\xcoord by -0.05\relax
      \raise\ycoord \Einheit\hbox to0pt{\hskip\xcoord \Einheit
         \unitlength\Einheit
         \line(1,-1){#3}\hss}}
\def\Pfad(#1,#2),#3\endPfad{\unskip\leavevmode
  \xcoord#1 \ycoord#2 \thicklines\ZeichnePfad#3\endPfad\thinlines}
\def\ZeichnePfad#1{\ifx#1\endPfad\let\next\relax
  \else\let\next\ZeichnePfad
    \ifnum#1=1
      \raise\ycoord \Einheit\hbox to0pt{\hskip\xcoord \Einheit
         \vrule height\Pfadd@cke width1 \Einheit depth\Pfadd@cke\hss}%
      \advance\xcoord by 1
    \else\ifnum#1=2
      \raise\ycoord \Einheit\hbox to0pt{\hskip\xcoord \Einheit
        \hbox{\hskip-\PfadD@cke\vrule height1 \Einheit width\PfadD@cke depth0pt}\hss}%
      \advance\ycoord by 1
    \else\ifnum#1=3
      \raise\ycoord \Einheit\hbox to0pt{\hskip\xcoord \Einheit
         \unitlength\Einheit
         \line(1,1){1}\hss}
      \advance\xcoord by 1
      \advance\ycoord by 1
    \else\ifnum#1=4
      \raise\ycoord \Einheit\hbox to0pt{\hskip\xcoord \Einheit
         \unitlength\Einheit
         \line(1,-1){1}\hss}
      \advance\xcoord by 1
      \advance\ycoord by -1
    \fi\fi\fi\fi
  \fi\next}
\def\hSSchritt{\leavevmode\raise-.4pt\hbox to0pt{\hss.\hss}\hskip.2\Einheit
  \raise-.4pt\hbox to0pt{\hss.\hss}\hskip.2\Einheit
  \raise-.4pt\hbox to0pt{\hss.\hss}\hskip.2\Einheit
  \raise-.4pt\hbox to0pt{\hss.\hss}\hskip.2\Einheit
  \raise-.4pt\hbox to0pt{\hss.\hss}\hskip.2\Einheit}
\def\vSSchritt{\vbox{\baselineskip.2\Einheit\lineskiplimit0pt
\hbox{.}\hbox{.}\hbox{.}\hbox{.}\hbox{.}}}
\def\DSSchritt{\leavevmode\raise-.4pt\hbox to0pt{%
  \hbox to0pt{\hss.\hss}\hskip.2\Einheit
  \raise.2\Einheit\hbox to0pt{\hss.\hss}\hskip.2\Einheit
  \raise.4\Einheit\hbox to0pt{\hss.\hss}\hskip.2\Einheit
  \raise.6\Einheit\hbox to0pt{\hss.\hss}\hskip.2\Einheit
  \raise.8\Einheit\hbox to0pt{\hss.\hss}\hss}}
\def\dSSchritt{\leavevmode\raise-.4pt\hbox to0pt{%
  \hbox to0pt{\hss.\hss}\hskip.2\Einheit
  \raise-.2\Einheit\hbox to0pt{\hss.\hss}\hskip.2\Einheit
  \raise-.4\Einheit\hbox to0pt{\hss.\hss}\hskip.2\Einheit
  \raise-.6\Einheit\hbox to0pt{\hss.\hss}\hskip.2\Einheit
  \raise-.8\Einheit\hbox to0pt{\hss.\hss}\hss}}
\def\SPfad(#1,#2),#3\endSPfad{\unskip\leavevmode
  \xcoord#1 \ycoord#2 \ZeichneSPfad#3\endSPfad}
\def\ZeichneSPfad#1{\ifx#1\endSPfad\let\next\relax
  \else\let\next\ZeichneSPfad
    \ifnum#1=1
      \raise\ycoord \Einheit\hbox to0pt{\hskip\xcoord \Einheit
         \hSSchritt\hss}%
      \advance\xcoord by 1
    \else\ifnum#1=2
      \raise\ycoord \Einheit\hbox to0pt{\hskip\xcoord \Einheit
        \hbox{\hskip-2pt \vSSchritt}\hss}%
      \advance\ycoord by 1
    \else\ifnum#1=3
      \raise\ycoord \Einheit\hbox to0pt{\hskip\xcoord \Einheit
         \DSSchritt\hss}
      \advance\xcoord by 1
      \advance\ycoord by 1
    \else\ifnum#1=4
      \raise\ycoord \Einheit\hbox to0pt{\hskip\xcoord \Einheit
         \dSSchritt\hss}
      \advance\xcoord by 1
      \advance\ycoord by -1
    \fi\fi\fi\fi
  \fi\next}
\def\Koordinatenachsen(#1,#2){\unskip
 \hbox to0pt{\hskip-.5pt\vrule height#2 \Einheit width.5pt depth1 \Einheit}%
 \hbox to0pt{\hskip-1 \Einheit \xcoord#1 \advance\xcoord by1
    \vrule height0.25pt width\xcoord \Einheit depth0.25pt\hss}}
\def\Koordinatenachsen(#1,#2)(#3,#4){\unskip
 \hbox to0pt{\hskip-.5pt \ycoord-#4 \advance\ycoord by1
    \vrule height#2 \Einheit width.5pt depth\ycoord \Einheit}%
 \hbox to0pt{\hskip-1 \Einheit \hskip#3\Einheit 
    \xcoord#1 \advance\xcoord by1 \advance\xcoord by-#3 
    \vrule height0.25pt width\xcoord \Einheit depth0.25pt\hss}}
\def\Gitter(#1,#2){\unskip \xcoord0 \ycoord0 \leavevmode
  \LOOP\ifnum\ycoord<#2
    \loop\ifnum\xcoord<#1
      \raise\ycoord \Einheit\hbox to0pt{\hskip\xcoord \Einheit\Punkt\hss}%
      \advance\xcoord by1
    \repeat
    \xcoord0
    \advance\ycoord by1
  \REPEAT}
\def\Gitter(#1,#2)(#3,#4){\unskip \xcoord#3 \ycoord#4 \leavevmode
  \LOOP\ifnum\ycoord<#2
    \loop\ifnum\xcoord<#1
      \raise\ycoord \Einheit\hbox to0pt{\hskip\xcoord \Einheit\Punkt\hss}%
      \advance\xcoord by1
    \repeat
    \xcoord#3
    \advance\ycoord by1
  \REPEAT}
\def\Label#1#2(#3,#4){\unskip \xdim#3 \Einheit \ydim#4 \Einheit
  \def\lo{\advance\xdim by-.5 \Einheit \advance\ydim by.5 \Einheit}%
  \def\llo{\advance\xdim by-.25cm \advance\ydim by.5 \Einheit}%
  \def\loo{\advance\xdim by-.5 \Einheit \advance\ydim by.25cm}%
  \def\o{\advance\ydim by.25cm}%
  \def\ro{\advance\xdim by.5 \Einheit \advance\ydim by.5 \Einheit}%
  \def\rro{\advance\xdim by.25cm \advance\ydim by.5 \Einheit}%
  \def\roo{\advance\xdim by.5 \Einheit \advance\ydim by.25cm}%
  \def\l{\advance\xdim by-.30cm}%
  \def\r{\advance\xdim by.30cm}%
  \def\lu{\advance\xdim by-.5 \Einheit \advance\ydim by-.6 \Einheit}%
  \def\llu{\advance\xdim by-.25cm \advance\ydim by-.6 \Einheit}%
  \def\luu{\advance\xdim by-.5 \Einheit \advance\ydim by-.30cm}%
  \def\u{\advance\ydim by-.30cm}%
  \def\ru{\advance\xdim by.5 \Einheit \advance\ydim by-.6 \Einheit}%
  \def\rru{\advance\xdim by.25cm \advance\ydim by-.6 \Einheit}%
  \def\ruu{\advance\xdim by.5 \Einheit \advance\ydim by-.30cm}%
  #1\raise\ydim\hbox to0pt{\hskip\xdim
     \vbox to0pt{\vss\hbox to0pt{\hss$#2$\hss}\vss}\hss}%
}
\catcode`\@=12

\def\al{\alpha}

\def\de{\delta}
\def\ep{\varepsilon}
\def\ze{\zeta}

\def\th{\theta}

\def\la{\lambda}
\def\rh{\rho}

\def\ph{\varphi}

\def\Ga{\Gamma}

\def\Th{\Theta}

\def\Z{{\mathbb Z}}

\def\today{\ifcase\month\or
 January\or February\or March\or April\or May\or June\or
 July\or August\or September\or October\or November\or December\fi
 \space\number\day, \number\year}

\def\({\left(}
\def\){\right)}
\def\[{\left[}
\def\]{\right]}

\def\Tr{\operatorname{Tr}}

\def\Re{\operatorname{Re}}

\def\3{\ss}

\input psfig.sty

\numberwithin{equation}{section}

\newcounter{saveeqn}
\newcommand{\alphaeqn}{\setcounter{saveeqn}{\value{equation}}%
\setcounter{equation}{0}%
\global\def\theequation{\mbox{\thesection.\arabic{saveeqn}\alph{equation}}}}
\newcommand{\reseteqn}{\setcounter{equation}{\value{saveeqn}}%
\global\def\theequation{\thesection.\arabic{equation}}}

\newtheorem{theorem}{Theorem}
\newtheorem{corollary}[theorem]{Corollary}
\newtheorem{lemma}[theorem]{Lemma} 
\newtheorem*{remark}{Remark}

\def\v#1{{\vert #1\vert}}
\def\Tr{\operatorname{Tr}}
\def\diag{\operatorname{diag}}
\def\po#1#2{(#1)_#2}
\def\fl#1{\left\lfloor#1\right\rfloor}

\def\bX{{\bar X}}

\begin{document}
\abovedisplayskip=5.9pt plus2pt minus 4pt
\belowdisplayskip=5.9pt plus2pt minus 4pt

\author[Christian Krattenthaler and Paul B. Slater]
{Christian Krattenthaler$^\dagger$ and Paul B. Slater}
\address{Institut f\"ur Mathematik\\
Universit\"at Wien\\
Strudlhofgasse 4\\
A-1090 Vienna, Austria\\
e-mail: {\tt kratt@pap.univie.ac.at}\\
WWW: {\tt http://radon.mat.univie.ac.at/People/kratt}}
\address{Institute for Social, Behavioral and Economic Research\\
University of California\\
Santa Barbara\\
CA 93106-2150\\
e-mail: slater@itp.ucsb.edu}
\title[Asymptotic Redundancies]{Asymptotic Redundancies for Universal
Quantum Coding}
{\def\thefootnote{}
\footnote{$^\dagger$Research supported in part by
the MSRI, Berkeley}}
\begin{abstract}
Clarke and Barron have recently shown that the Jeffreys' invariant
prior of Bayesian theory yields the common asymptotic
(minimax and maximin) redundancy of universal data compression
in a parametric setting. We seek  a possible analogue
of this result for the two-level {\it quantum} systems.
We restrict our considerations to
 prior
probability distributions belonging
to a certain
 one-parameter family, $q_{u}$, $-\infty < u < 1$.
Within this setting, we 
are able to compute exact redundancy formulas, for which we find
 the asymptotic limits.
We compare our
quantum asymptotic redundancy formulas to those derived by
naively applying the classical counterparts
 of Clarke and Barron, and find certain common features.
Our results are based on formulas we obtain for the
eigenvalues and eigenvectors of  $2^n \times 2^n$ (Bayesian 
density) matrices, $\zeta_{n}(u)$.
These matrices are the weighted averages (with respect to $q_{u}$)
of all possible tensor products of $n$ identical $2 \times 2$ density matrices,
representing the two-level quantum systems.
We propose a form of {\it universal} coding 
for the situation in which
 the density matrix describing an ensemble of quantum signal states
is unknown. A sequence of $n$ signals would be
projected onto the dominant eigenspaces of $\ze_n(u)$.

\end{abstract}

\maketitle

\vspace{48pt} 
\section{Introduction}\label{s1}

In recent years, there have been a considerable number
of important developments in the
extension of (classical) information-theoretic concepts to a 
quantum-mechanical setting. Bennett and Shor \cite{bennett} have surveyed
this progress in the outstanding Commemorative Issue 1948--1998 of the
{\it IEEE Transactions on Information Theory}.
In particular, they pointed out --- in strict analogy to the
classical case, successfully studied some fifty years ago by Shannon 
in famous landmark work \cite{shann} --- that
quantum data compression allows signals from a redundant quantum source
to be compressed into a bulk approaching the source's (quantum) entropy.
Bennett and Shor did not, however, discuss the 
intriguing case which arises when the specific nature of the quantum
source is {\it unknown}. This, of course,
corresponds to the classical question of {\it universal} coding
or data compression (see \cite{davi}, \cite[Sec.~II.E]{verdu}).

We do address this interesting  
issue here, by investigating whether or not it is
possible to extend to the quantum domain,
recent (classical) seminal results of Clarke and Barron \cite{cl3,cl1,cl2}.
They, in fact, derived various forms of asymptotic redundancy of universal
data compression for parameterized families of
probability distributions. Their analyses
provide a rigorous basis for the reference prior method in Bayesian
statistical analysis.
For an extensive commentary on the
results of Clarke and Barron, see \cite{ri}. Also see \cite{cl4}, for
some recent related research, as well as a
discussion of various rationales
that have been employed for using the (classical) Jeffreys'
 prior --- a possible
quantum counterpart of which will be of interest here --- for
Bayesian purposes, cf.~\cite{kass}.
Let us also bring to the attention
of the reader that in a brief review 
\cite{good1} of \cite{cl1}, the noted
statistician, I. J. Good, commented that Clarke and Barron ``have presumably
overlooked the reviewer's work'' and cited, in this regard
 \cite{good2,good3}.%
\footnote{It should be noted that in these papers, Good uses a more general
objective function --- a two-parameter utility --- than the relative
entropy, chosen by Clarke and Barron over
alternative measures \cite[p.~454]{cl3}. Good does conclude that Jeffreys'
invariant prior is the minimax, that is, the least favorable, prior when
the utility is the ``weight of the evidence'' in the sense of C. S.
Pierce, that is, the relative entropy.}

Let us briefly recall the basic setup and the results of Clarke and
Barron that are relevant to the analyses of our paper. Clarke and
Barron work in a noninformative Bayesian framework, in which we are
given a parametric family of probability densities
$\{P_\th:\th\in\Th\subseteq \mathbb R^d\}$ on a space $X$. These
probability densities generate independent identically distributed
random variables $X_1,X_2,\dots,X_n$, 
which, for a fixed $\th$, we consider as producing
strings of length $n$ according to the probability density
$P_\th^n$ of the $n$-fold product of probability
distributions. Now suppose
that Nature picks a $\th$ from $\Th$, that is a joint density
$P_\th^n$ on the product space $X^n=(X_1,X_2,\dots,X_n)$, 
the space of strings of length $n$. On the other
hand, a Statistician chooses a distribution $Q_n$ on $X^n$ as his
best guess of $P_\th^n$. Of course, there is a loss of information,
which is measured by the total relative entropy $D(P_\th^n\Vert
Q_n)$, where $D(P\Vert Q)$ is the 
{\it Kullback--Leibler divergence} of $P$
and $Q$ (the {\it relative entropy} of $P$ with respect to $Q$). 
For finite $n$, and for a given {\it prior}
$w(\th)d\th$ on $\Th$, by a result of Aitchison \cite[pp.~549/550]{ait}, 
the best strategy $Q_n$ to minimize the average risk $\int
D(P_\th^n\Vert Q_n)w(\th)\,d\th$ is 
to choose for $Q_n$ the mixture density $M_n^w=\int
P_\th^nw(\th)\,d\th$. 
This is called a {\it Bayes procedure} or a {\it Bayes strategy}.

The quantities corresponding to such a procedure that must be
investigated are the 
{\it risk} ({\it redundancy}) {\it 
of the Bayes strategy} $D(P_\th^n\Vert
M_n^w)$ and the {\it Bayes risk}, 
the average of risks, $\int D(P_\th^n\Vert
M_n^w)w(\th)\,d\th$. The Bayes risk equals Shannon's mutual
information $I(\Th;X^n)$ (see \cite{cl3,davi}). Moreover,
the Bayes risk is bounded above by the {\it minimax
redundancy} $\min_{Q_n}\max_{\th\in\Th}D(P_\th^n\Vert Q_n)$. In fact, 
by a result of Gallager \cite{gall} and Davisson and Leon--Garcia 
\cite{davi-garcia} (see \cite{hauss} for a generalization), for
each fixed $n$ there is a prior $w_n^*$ which realizes this upper
bound, i.e., the {\it maximin redundancy} $\max_w\int D(P_\th^n\Vert
M_n^w)w(\th)\,d\th$ and the minimax redundancy are the same. Such a prior
$w_n^*$ is called {\it capacity achieving} or {\it least favorable}.

Clarke and Barron investigate the above-mentioned quantities
{\it asymptotically}, that is, for $n$ tending to infinity. First of all,
in \cite[(1.4)]{cl3}, \cite[(2.1b)]{cl1},
they show that the redundancy $D(P_\th^n\Vert M_n^w)$ 
of the Bayes strategy is asymptotically
\begin{equation} \label{eq:4}
{\frac d  2} {\log {\frac n  {2 \pi e}}}+{\frac 1  2}\log \det
I(\theta)-\log w(\theta) +o(1),
\end{equation}
as $n$ tends to infinity. Here, $I(\theta)$
is the $d \times d$ Fisher information matrix --- the
negative of the expected value of the Hessian
of the logarithm of the density function.
(Although the binary logarithm is usually used in
the quantum coding literature, we employ the natural logarithm
throughout this paper, chiefly to facilitate comparisons of
our results with those of Clarke and Barron \cite{cl3,cl1,cl2}.)
For priors supported on a compact subset $K$ in the interior of the
domain $\Th$ of parameters, 
the asymptotic minimax redundancy 
$\min_{Q_n}\max_{\th\in\Th}D(P_\th^n\Vert Q_n)$
was shown to be \cite[(2.4)]{cl1},
\cite{cl2},
\begin{equation} \label{eq:3}
{\frac d  2} {\log {\frac n  {2 \pi e}}} +\log {\int_{K} \sqrt{\det
I(\theta)}} \,
d \theta + o(1).
\end{equation}
Moreover \cite[(2.6)]{cl1}, 
it is {\it Jeffreys' prior} $w^*=\sqrt{\det I(\th)}/c$ (with
$c=\int_K \sqrt{\det I(\th)}$ a normalizing constant; 
see also
\cite{bern}) which is
the unique continuous and positive prior on $K$ which is
asymptotically least favorable, i.e., for which the asymptotic
maximin redundancy achieves the value (\ref{eq:3}). In particular,
asymptotically the maximin and minimax redundancies are the same.

\medskip
In obvious contrast to classical information theory, quantum information theory
directly relies upon the fundamental principles of quantum mechanics. This is 
due to the fact that
the basic unit of quantum computing,
the ``quantum bit" or ``qubit,'' is typically a (two-state) microscopic system, 
possibly an atom or
nuclear spin or polarized photon, the behavior of which (e.g\@.
entanglement, interference, superposition, stochasticity, \ldots) can only be
accurately explained using the rules of quantum theory
\cite{per}. We refer the reader to
 \cite{bennett} for
a comprehensive introduction to these matters
(including the subjects of quantum error-correcting codes and quantum
cryptography). Here, we shall restrict
ourselves to describing, in mathematical terms, the basic notions of quantum
information theory, how they pertain to
data compression, and in what manner they parallel the corresponding notions
from classical information theory.

In quantum information theory, the role of probability 
densities is played by
{\it density matrices}, which are, by definition, nonnegative definite 
Hermitian matrices of unit trace, and which can be considered as
operators acting on a (finite-dimensional) Hilbert space. Any probability
density on a (finite) set $X=\{x_1,x_2,\dots,x_m\}$, where the
probability of $x_i$ equals $p_i$, is representable in  this framework
by a diagonal matrix $\diag(p_1,p_2,\dots,p_m)$ (which is quite clearly
itself, a 
nonnegative definite Hermitian matrix with unit trace).
Given two density matrices $\rh_1$ and $\rh_2$, the quantum counterpart
of the relative entropy, that is, 
the {\it relative entropy} of $\rho_{1}$ with respect
to $\rho_{2}$, is \cite{oh,wehrl} (cf.~\cite{pe4}),
\begin{equation} \label{eq:5}
 S(\rho_{1},\rho_{2}) =
 \Tr \rho_{1} (\log \rho_{1} -  \log \rho_{2}),
\end{equation}
where the logarithm of a matrix $\rh$ is defined as $\sum _{k\ge1}
^{}(-1)^{k-1}(\rh-I)^k/k$, with $I$ the appropriate identity matrix.
(Alternatively, if $\rho$ acts diagonally on a basis
$\{v_1,v_2,\dots,v_m\}$ of the Hilbert space by $\rho v_i=\la_iv_i$,
then $\log\rho$ acts by $(\log\rho) v_i=(\log\la_i)v_i$,
$i=1,2,\dots,m$.)
Clearly, if $\rh_1$ and $\rh_2$ are diagonal matrices, corresponding
to classical probability densities, then (\ref{eq:5}) reduces to the
usual Kullback--Leibler divergence.

As we said earlier, our goal is to examine the possibility of
extending the results of Clarke and Barron to quantum theory. That
is, first of all we have to replace the (classical) probability
densities $P_\th$ by density matrices. We are not able to proceed in
complete generality, but rather we will restrict ourselves to
considering the first nontrivial case, that is, we will replace 
$P_\th$ by $2\times 2$ density matrices. Such matrices can be written 
in the form,
\begin{equation} \label{eq:6}
\rho = {\frac 1  2} \begin{pmatrix}1+z&x-iy\\
                   x+iy&1-z\end{pmatrix},
\end{equation}
where, in order to guarantee
nonnegative definiteness, the points $(x,y,z)$ must lie within
the unit ball (``Bloch sphere'' \cite{br}), $x^2 +y^2+z^2 \leq 1$.
(The points on the bounding spherical surface, $x^2+y^2+z^2=1$,
corresponding to the {\it pure states}, will be shown 
to exhibit nongeneric behavior, 
see (\ref{a5}) and the respective comments in Sec.~\ref{s3}
(cf.~\cite{fuj}).)
Such $2\times 2$ density matrices correspond, in a one-to-one fashion, to the 
standard 
(complex) two-level quantum systems --- notably, those of
spin-$1/2$ (electrons, protons,\dots)
and massless spin-$1$ particles (photons). These systems carry the
basic units of quantum computing, the {\it quantum bits}.
(If we set $x=y=0$ in (\ref{eq:6}), we recover a
classical binomial distribution, with the probability of ``success'', say, 
being $(1+z)/2$ and of ``failure'', $(1-z)/2$. Setting either
$x$ or $y$ to zero, puts us in the
framework of real --- as opposed to complex --- quantum
mechanics.) 

The quantum analogue of the product of (classical) probability
distributions
is the {\it tensor product\/} of density matrices. (Again, it is easily seen
that, for diagonal matrices, this reduces to the classical product.)
Hence, we will replace $P_\th^n$ by the tensor products $\overset n  
\otimes \rho$, where $\rho$ is a $2\times 2$ density matrix
(\ref{eq:6}). These tensor products are $2^n\times 2^n$ matrices, and
can be used to compute ({\it via} the fundamental rule that the expected value
of an observable is the trace of the matrix product of the observable
and the density matrix; see \cite{per})
the probability of strings of quantum bits of length $n$.

In \cite{sl1,sl2}  it was argued that the quantum Fisher information
matrix (requiring --- due to noncommutativity ---
the computation of symmetric logarithmic
derivatives 
\cite{pe1}\footnote{Following \cite{fuj},
in order to derive (\ref{eq:8}), one must find the symmetric 
logarithmic derivatives ($L_{x},L_{y},L_{z}$) satisfying
\begin{equation}\notag
\frac{\partial \rho } { \partial \alpha} = 
(\rho L_{\alpha} + L_{\alpha} \rho)/ 2,\quad \alpha = x, y, z,
\end{equation}
and then compute the  entries of (\ref{eq:8}) in the form
\cite[eqs. (2), (3)]{sl1}
\begin{equation}\notag
I_{\beta \gamma} =\Tr[\rho (L_{\beta} L_{\gamma} + L_{\gamma}
L_{\beta})/ 2], \quad \beta, \gamma = x, y, z.
\end{equation}
For a well-motivated discussion of these formulas and the manner
in which classical and quantum Fisher information are related,
see \cite{malley}.}) for the density matrices (\ref{eq:6}) should
be taken to be of the form
\begin{equation} \label{eq:8}
{\frac1 {(1-x^2-y^2-z^2)}} \begin{pmatrix}1-y^2-z^2&xy&xz\\
                                 xy&1-x^2-z^2&yz\\
                                 xz&yz&1-x^2-y^2\end{pmatrix}\quad.
\end{equation}
The quantum counterpart of the
Jeffreys' prior was, then, taken to be the normalized form
(dividing by $\pi^2$)
of the square root of the determinant of (\ref{eq:8}), that is,
\begin{equation} \label{eq:9}
 (1-x^2-y^2-z^2)^{-1/2}/{\pi}^2.
\end{equation}

On the basis of the above-mentioned result of Clarke and Barron 
that the Jeffreys' prior yields the asymptotic common minimax
and maximin redundancy, it was conjectured
\cite{sl4} that its assumed quantum counterpart (\ref{eq:9}) would
have similar properties, as well.

To examine this possibility,
(\ref{eq:9}) was embedded as a specific member ($u=.5$) of a
one-parameter family of 
spherically-symmetric/unitarily-invariant probability 
densities 
(i.e., under unitary transformations of $\rho$, 
the assigned probability is invariant),
\begin{equation} \label{eq:10}
q_{u}=q_{u}(x,y,z):=\frac {\Gamma(5/2-u)} 
{\pi^{3/2}\,\Gamma(1-u)\,(1-x^2-y^2-z^2)^u},\quad -\infty<u<1 .
\end{equation}
(Embeddings of (\ref{eq:9}) in other (possibly, multiparameter)
families are, of course, possible and may be pursued
in further research. 
In this regard, see Theorem~\ref{t15} in Sec.~\ref{s3}.)
For $u=0$, we obtain a uniform distribution over the unit ball.
(This has been used as a prior over the two-level quantum systems,
at least, in one study \cite{larson}.) For $u \rightarrow 1$,
the uniform distribution over the spherical boundary
(the locus of the pure states) is approached.
(This is often employed as a prior, for example \cite{jones,larson,massar}.)
For $u \rightarrow -\infty$, a Dirac distribution concentrated at
the origin (corresponding to the fully mixed state) is approached.

For a treatment in our setting that is analogous to that of Clarke
and Barron, we average $\overset n\otimes \rho$ with respect to
$q_{u}$. Doing so yields a one-parameter family 
of $2^n \times 2^n$ {\it Bayesian density matrices}
\cite{cl2,cl3,mat}, 
$$\zeta_{n}(u)=\int_{x^2+y^2+z^2\le 1}\big(\overset n\otimes 
\rho \big)q_{u}(x,y,z)\,dx\,dy\,dz,$$
$-\infty < u < 1$, which are the analogues of the mixtures $M_n^w$,
and which exhibit highly interesting properties.

Now, still following Clarke and Barron, we have to compute the
analogue of the risk $D(P_\th^n\Vert M_n^w)$, i.e., the relative
entropy $S(\overset n\otimes 
\rho ,\ze_n(u))$. Keeping the  definition (\ref{eq:5}) in mind, this
requires us to explicitly find  
the eigenvalues and eigenvectors of the matrices
$\ze_n(u)$, which we do in Sec.~\ref{s2.2}. 
Subsequently, in Sec.~\ref{s2.3}, we 
determine explicitly the relative entropy of
$\overset n\otimes \rho$ with respect to $\ze_n(u)$. 
We do this by using identities for hypergeometric series and some
combinatorics. (It is also possible to obtain some of our results by
making use of representation theory of $SU(2)$. An even more general
result was derived by combining these two approaches. We comment on
this issue at the end of Sec.~\ref{s3}.)

On the basis of these results, we then address the question of
finding asymptotic estimations in Sec.~\ref{s2.4} and \ref{s2.5}.
These, in turn, form the basis of examining to what degree the results
of Clarke and Barron are capable of extension to the quantum domain.

Let us (naively) attempt to apply the formulas
of Clarke and Barron \cite{cl1,cl2} --- (\ref{eq:4}) and (\ref{eq:3})
above --- to
the quantum context under investigation here. We do this by
setting $d$ to 3 (the dimensionality of the unit ball --- which we
take as $K$), $\det I(\theta)$ to $(1-x^2-y^2-z^2)^{-1}$ 
(the determinant of the quantum Fisher information matrix (\ref{eq:8})),
so that $\int_{K} \sqrt {\det I(\theta)}\,d\th$ is $\pi^2$, and
$w(\theta)$ to $q_{u}(x,y,z)$. Then, we obtain from 
the expression for the asymptotic redundancy (\ref{eq:4}),
\begin{equation} \label{eq:12}
{\frac 3  2} (\log n - \log 2 -1) -\(\frac {1} {2}-u\) \log (1-r^2) + \log \Gamma 
(1-u)
- \log \Gamma\({\frac 5  2} - u\) + o(1),
\end{equation}
where $r=\sqrt{x^2+y^2+z^2}$,
and from the expression for the asymptotic minimax redundancy (\ref{eq:3}),
\begin{equation} \label{eq:11}
{\frac 3  2} (\log n -\log 2 -1) + {\frac 1  2} \log \pi + o(1).
\end{equation}

We shall
(in Sec.~\ref{s3})
compare these two formulas,
(\ref{eq:12}) and (\ref{eq:11}), with the results of Sec.~\ref{s2}
and find some striking similarities
and coincidences, particularly associated with the
fully mixed state ($r=0$).
These findings will help to support the working hypothesis
of this study --- that there are meaningful extensions to the
quantum domain of the 
(commutative probabilistic) theorems of Clarke and Barron.
However, we find that 
the minimax and maximin properties of
the Jeffreys' prior do not strictly carry over, 
but transfer only in an approximate sense, which is, nevertheless, still quite
remarkable.
In any case, we can not formally rule out the possibility that the
actual global (perhaps common) minimax and maximin are achieved for
probability distributions not belonging to the one-parameter
family $q_{u}$.

In analogy to \cite[Sec.~5.2]{cl1},
the matrices $\ze_n(u)$ should prove useful for 
the {\it universal} version of Schumacher data compression
\cite{ben,cleve,jo,sch}. 
Schumacher's result \cite{sch,jo} must be
considered as the quantum analogue of Shannon's noiseless coding theorem 
(see e.g\@. \cite[Sec.~5.6]{welsh}).
Roughly, {\it quantum data compression}, as proposed by
Schumacher \cite{sch}, works as follows:
A (quantum) signal source (``sender") generates signal states of a
quantum system $M$, the
ensemble of possible signals being described by a density operator
$\psi$. The signals are projected down to a ``dominant" subspace of
$M$, the rest is discarded. The information in this dominant subspace
is transmitted through a (quantum) channel. The receiver tries to reconstruct
the original signal by replacing the discarded information by some 
``typical" state. The quality (or {\it faithfulness}) of a coding scheme is
measured by the {\it fidelity}, which is by definition the overall
probability that a signal from the signal ensemble $M$ that is
transmitted to the receiver passes a validation test comparing it to
its original (see \cite[Sec.~IV]{sch}). 
What Schumacher shows is that, for each $\ep>0$ and $\de>0$, under
the above coding scheme
a compression rate of $S(\psi)+\de$ qubits per signal is possible,
where $S(\psi)$ is the {\it von Neumann entropy} of $\psi$,
\begin{equation} \label{eq:1}
S(\psi) = -\Tr \psi \log \psi,
\end{equation}
at a fidelity of 
at least $1-2\ep$. (Thus, the von Neumann entropy is the
quantum analogue of the Shannon entropy, which features in
Shannon's classical noiseless coding theorem. Indeed, as is easy
to see, for diagonal matrices, corresponding to classical probability
densities, the right-hand side of (\ref{eq:1}) reduces to the Shannon
entropy.) This is achieved by choosing as the dominant subspace that subspace
of the quantum system $M$ which is the span of the eigenvectors of
$\psi$ corresponding to the largest eigenvalues, with the property
that the eigenvalues add up to at least $1-\ep$.

Consequently,
in a universal compression scheme, we propose to
project blocks of $n$ signals (qubits) onto those
``typical'' subspaces 
of $2^n$-dimensional Hilbert space corresponding to as many of
the dominant eigenvalues of $\zeta_{n}(u)$ as it takes to exceed a
sum $1- \ep$. For all $u$, the leading 
one of the $\fl {\frac {n} {2}} + 1$ 
distinct eigenvalues has multiplicity $n+1$, and belongs to the
($n+1$)-dimensional (Bose--Einstein) symmetric subspace \cite{Ba1}.
(Projection onto the symmetric subspace has been proposed as a
method for stabilizing quantum computations, including quantum
state storage \cite{bar}.)
For $u=1/2$, the leading eigenvalue can be obtained by dividing
the $(n+1)$-st Catalan number 
--- that is, 
$\frac {1} {n+2}\binom {2(n+1)}{n+1}$ --- by $4^n$.
(The Catalan numbers ``are probably the most frequently occurring
combinatorial numbers after the binomial coefficients'' \cite{sloane}.)

Let us point out to the reader the
quite recent important work of Petz and Sud\'ar \cite{pe1}.
They demonstrated that in the quantum case --- in
 contrast to the classical situation
in which there is, as originally shown by Chentsov \cite{chen}, essentially only 
one monotone metric and,
therefore, essentially only one form of the Fisher information --- there
 exists an infinitude of such metrics.
``The monotonicity of the Riemannian metric $g$ is crucial when one likes to
imitate the geometrical approach of [Chentsov]. An infinitesimal statistical
 distance
has to be monotone under stochastic mappings. We note that the monotonicity
of $g$ is a strengthening of the concavity of the von Neumann entropy.
Indeed, positive definiteness of $g$ is equivalent to the strict concavity
of the von Neumann entropy
\dots{} and monotonicity is much more than positivity''
\cite{pe3}.

 The monotone metrics on the space of density matrices are
given \cite{pe1} by the operator monotone functions $f(t):
\mathbb R^+ \rightarrow
\mathbb R^+$, such that
$f(1)=1$ and $f(t) = t f(1/t)$. For the choice $f= (1+t)/2$,
one obtains the minimal metric (of the symmetric logarithmic
derivative), which serves as the basis of our analysis here.
``In accordance with the work of Braunstein and Caves, this seems
to be the canonical metric of parameter estimation theory. However,
expectation values of certain relevant observables are known to lead
to statistical inference theory provided by the maximum entropy
principle or the minimum relative entropy principle when {\it a priori}
information on the state is available. The best prediction is a kind of
generalized Gibbs state. On the manifold of those states, the
differentiation of the entropy functional yields the
 Kubo-Mori/Bogoliubov metric,
which is different from the metric of the symmetric logarithmic
derivative. Therefore, more than one privileged metric shows up
in quantum mechanics. The exact clarification of this point requires
and is worth further studies'' \cite{pe1}.
It remains a possibility, then, that a monotone metric other than the
minimal one (which corresponds to $q_{0.5}$, that is (\ref{eq:9})) may yield
a  common global asymptotic minimax and maximin
 redundancy, thus, fully paralleling
the classical/nonquantum results of Clarke and Barron
\cite{cl3,cl1,cl2}. We intend to investigate such a possibility,
in particular, for the Kubo-Mori/Bogoliubov metric \cite{pe3,pe1,pe2}.

\section{Analysis of a One-Parameter Family of Bayesian Density Matrices}
\label{s2}
In this section, we implement the analytical approach described in the
Introduction to extending the work of Clarke and Barron \cite{cl1,cl2}
to the realm of quantum mechanics, specifically, the two-level
systems. Such systems are representable by density matrices $\rho$ of the
form (\ref{eq:6}). A composite system of $n$ independent
(unentangled) and identical
two-level quantum systems is, then, represented by the $n$-fold tensor
product $\overset n\otimes \rho$. In
Theorem~\ref{t1} of Sec.~\ref{s2.1},
we average
$\overset n\otimes\rho$ with respect to the one-parameter family of 
probability 
densities $q_{u}$ defined in (\ref{eq:10}), obtaining
the Bayesian density matrices $\ze_n(u)$
and formulas for their $2^{2 n}$ entries. Then, in
Theorem~\ref{t2}
of Sec.~\ref{s2.2}, we are able to explicitly determine the $2^n$ eigenvalues
and eigenvectors of $\ze_n(u)$. Using these results, in
Sec.~\ref{s2.3}, we compute the relative entropy of $\overset
n\otimes\rho$ with respect to $\ze_n(u)$. Then, in Sec.~\ref{s2.4}, we
obtain the asymptotics of this relative entropy for $n\to\infty$. 
In Sec.~\ref{s2.5}, we compute the asymptotics of the von Neumann
entropy (see (\ref{eq:1})) of $\ze_n(u)$.
All these results 
will enable us, in Sec.~\ref{s3}, to ascertain to what extent the
results of Clarke and Barron
could be said to carry over to the quantum domain.

\subsection{Entries of the Bayesian density matrices
$\ze_n(u)$}\label{s2.1}
The $n$-fold tensor product $\overset n\otimes \rho$ is a
$2^n\times 2^n$ matrix. To refer to specific rows and columns of 
$\overset n\otimes \rho$, we index them by subsets of
the $n$-element set $\{1,2,\dots,n\}$. We choose to employ this
notation instead of the more familiar use of binary strings, in order 
to have a
more succinct way of writing our formulas. For convenience, 
we will subsequently write $[n]$ for $\{1,2,\dots,n\}$.
Thus, $\overset n\otimes \rho$ can be written in the form
$$\overset n\otimes \rho=\begin{pmatrix} R_{IJ}\end{pmatrix}_{I,J\in
[n]},$$
where
\begin{equation}\label{e2}
R_{IJ}=\frac {1} {2^n}(1+z)^{n_{\in\in}} (1-z)^{n_{\notin\notin}}
(x+iy)^{n_{\notin\in}} (x-iy)^{n_{\in\notin}},
\end{equation}
with $n_{\in\in}$ denoting the number of elements of $[n]$
contained in both $I$ and $J$, $n_{\notin\notin}$ denoting the number
of elements {\it not\/} in both $I$ and $J$, $n_{\notin\in}$ denoting the number
of elements not in $I$ but in $J$, and $n_{\in\notin}$ denoting the number
of elements in $I$ but not in $J$. In symbols,
\begin{align*} n_{\in\in}&=\v{I\cap J},\\ n_{\notin\notin}&=\v{[n]
\backslash (I\cup J)},\\ n_{\notin\in}&=\v{J\backslash I},\\
n_{\in\notin}&=\v{I\backslash J}.
\end{align*}
We consider the average $\ze_n(u)$ of $\overset n\otimes \rho$ with respect
to the probability 
density $q_{u}=q_u(x,y,z)$ defined in (\ref{eq:10})
taken over the unit sphere $\{(x,y,z):x^2+y^2+z^2\le 1\}$.
This average can be described explicitly as follows.
\begin{theorem}\label{t1}
The average $\ze_n(u)$,
$$\int_{x^2+y^2+z^2\le 1}\big(\overset n\otimes \rho\big)\,
q_{u}(x,y,z)\,dx\,dy\,dz,$$ 
equals the matrix $(Z_{IJ})_{I,J\in[n]}$,
where
\begin{multline} \label{e4}
Z_{IJ}=\de_{n_{\notin\in},n_{\in\notin}}\big(\tfrac {n-n_{\in\in}-
n_{\notin\notin}} 2\big)!\,\\
\times
\frac {1} {2^n}\frac {\Ga\(\frac {5} {2}-u\)\,\Ga\(2+\frac {n} {2}+
\frac{n_{\in\in}}2-\frac {n_{\notin\notin}}2-u\)\, 
\Ga\(2+\frac {n} {2}+\frac {n_{\notin\notin}}2-
\frac {n_{\in\in}}2-u\)} {\Ga\(\frac {5} {2}+\frac {n} {2}-u\)\,
\Ga\(2+\frac {n} {2}-u\)\, \Ga\(2+\frac {n} {2}-\frac {n_{\in\in}}2-
\frac {n_{\notin\notin}}2-u\)}.
\end{multline}
Here, $\de_{i,j}$ denotes the Kronecker delta, $\de_{i,j}=1$ if $i=j$
and $\de_{i,j}=0$ otherwise.
\end{theorem}

\begin{remark} \em It is important for later considerations to observe
that because of the term $\de_{n_{\notin\in},n_{\in\notin}}$ in (\ref{e4})
the entry $Z_{IJ}$ is nonzero if and only if the sets $I$ and $J$
have the same cardinality. If $I$ and $J$ have the same cardinality,
$c$ say,
then $Z_{IJ}$ only depends on $n_{\in\in}$, the number of common
elements of $I$ and $J$, since in this case $n_{\notin\notin}$ is
expressible as $n-2c+n_{\in\in}$.
\end{remark}
\medskip
{\sc Proof of Theorem~\ref{t1}}.
To compute $Z_{IJ}$, we have to compute the integral 
\begin{equation}\label{e5}
\int_{x^2+y^2+z^2\le 1}R_{IJ}\,q_{u}(x,y,z)\,dx\,dy\,dz.
\end{equation}
For convenience, we treat the case that $n_{\in\in}\ge n_{\notin\notin}$
and $n_{\notin\in}\ge n_{\in\notin}$. The other four cases are
treated similarly.

First, we rewrite the matrix entries $R_{IJ}$,
\begin{align} \notag
\frac {1} {2^n}(1+z)^{n_{\in\in}}& (1-z)^{n_{\notin\notin}}
(x+iy)^{n_{\notin\in}} (x-iy)^{n_{\in\notin}}\\
\notag
=&\frac {1} {2^n}(1-z^2)^{n_{\in\in}}
(1-z)^{n_{\notin\notin}-n_{\in\in}}
(x^2+y^2)^{n_{\notin\in}} (x-iy)^{n_{\in\notin}-n_{\notin\in}}\\
\notag
=&\frac {1} {2^n}\sum _{j,k,l\ge0} ^{}(-1)^{j+k}(-i)^l
\binom {n_{\in\in}}j \binom
{n_{\notin\notin}-n_{\in\in}}k \binom
{n_{\in\notin}-n_{\notin\in}}l \\
&\hskip2cm\cdot z^{2j+k}(x^2+y^2)^{n_{\notin\in}}
x^{n_{\in\notin}-n_{\notin\in}-l} y^l.\label{e6}
\end{align}
Of course, in order to compute the integral (\ref{e5}), we transform the
Cartesian coordinates into polar coordinates,
\begin{align}\notag x&=r\sin\vartheta \cos\ph\\
y&=r\sin\vartheta \sin\ph\notag\\
z&=r\cos\vartheta,\notag\\
0\le \ph\le {}&2\pi,\ 0\le\vartheta\le\pi.\notag
\end{align}
Thus, using (\ref{e6}), the integral (\ref{e5}) is transformed into
\begin{multline} \label{e7}
\frac {1} {2^n}\sum _{j,k,l\ge0} ^{}
\int_0^1\int_0^{\pi}\int_0^{2\pi} (-1)^{j+k}(-i)^l
\binom {n_{\in\in}}j \binom
{n_{\notin\notin}-n_{\in\in}}k \binom
{n_{\in\notin}-n_{\notin\in}}l \\
\hskip2cm\cdot r^{2j+k+n_{\notin\in}+n_{\in\notin}+2}
\(\cos^{2j+k}\vartheta\)\(\sin^{n_{\notin\in}+n_{\in\notin}+1}\vartheta \)\\
\hskip2cm\cdot \(\cos^{n_{\in\notin}-n_{\notin\in}-l}\ph \)\(\sin^l\ph\)
\frac {\Ga(5/2-u)} {\pi^{3/2}\,\Ga(1-u)\,(1-r^2)^u}
\,d\ph\,d\vartheta\,dr.
\end{multline}
To evaluate this triple integral we use the following standard
formulas:
{\refstepcounter{equation}\label{e8}}
\alphaeqn
\begin{align} \int_0^\pi\sin^{2M}\vartheta\,
\cos^{2N}\vartheta\,d\vartheta&=\pi\frac {(2M-1)!!\,(2N-1)!!}
{(2M+2N)!!},\label{e8a}\\
\int_0^\pi\sin^{2M+1}\vartheta\,
\cos^{2N}\vartheta\,d\vartheta&=2\frac {(2M)!!\,(2N-1)!!}
{(2M+2N+1)!!}\ \\
&\text {and}\quad 
\int_0^{2\pi}\sin^{2M+1}\vartheta\,
\cos^{2N}\vartheta\,d\vartheta=0,\label{e8b}\\
\int_0^\pi\sin^{2M}\vartheta\,
\cos^{2N+1}\vartheta\,d\vartheta&=0,\label{e8c}\\
\int_0^\pi\sin^{2M+1}\vartheta\,
\cos^{2N+1}\vartheta\,d\vartheta&=0,\label{e8d}
\end{align}
\reseteqn
for any nonnegative integers $M$ and $N$. Furthermore, we need
the beta integral
\begin{equation}\label{e9}
\int_0^1\frac {r^m} {(1-r^2)^u}\,dr=\frac {\Ga\( \frac {m+1}
{2}\)\,\Ga(1-u)} {2\,\Ga\(\frac {m+3} {2}-u\)}.
\end{equation}

Now we consider the integral over $\ph$ in (\ref{e7}). Using
(\ref{e8b}) and (\ref{e8c}),
we see that each summand in (\ref{e7}) vanishes if $n_{\notin\in}$ has a
parity different from $n_{\in\notin}$. On the other hand, if 
$n_{\notin\in}$ has the same parity as $n_{\in\notin}$, then we can
evaluate the integrals over $\ph$ using (\ref{e8a}) and
(\ref{e8d}). 
Discarding for a
moment the terms independent of $\ph$ and $l$, we have
\begin{align} \sum _{l\ge0} ^{}&\int_0^{2\pi} (-i)^l  \binom
{n_{\in\notin}-n_{\notin\in}}l 
\(\cos^{n_{\in\notin}-n_{\notin\in}-l}\ph \)\(\sin^l\ph\)\,d\ph\notag\\
&=\sum _{l\ge0} ^{}(-1)^l\binom
{n_{\in\notin}-n_{\notin\in}}{2l} 2\pi\,\frac {(2l-1)!!\,(n_{\in\notin}
-n_{\notin\in}-2l-1)!!} {(n_{\in\notin}-n_{\notin\in})!!}\notag\\
&=2\pi\frac {(n_{\in\notin}-n_{\notin\in}-1)!!} 
{(n_{\in\notin}-n_{\notin\in})!!}\sum _{l\ge0} ^{}\binom 
{(n_{\in\notin}-n_{\notin\in})/2}l
(-1)^l\notag\\
&=2\pi\,\de_{n_{\in\notin},n_{\notin\in}},\notag
\end{align}
the last line being due to the binomial theorem. These 
considerations reduce (\ref{e7}) to
\begin{multline}\notag \de_{n_{\in\notin},n_{\notin\in}}\frac {1} {2^n}
\sum _{j,k\ge0} ^{}
\int_0^1\int_0^{\pi} (-1)^{j+k}
\binom {n_{\in\in}}j \binom
{n_{\notin\notin}-n_{\in\in}}k \\
\hskip2cm\cdot r^{2j+k+2n_{\notin\in}+2}
\(\cos^{2j+k}\vartheta\)\(\sin^{2n_{\notin\in}+1}\vartheta \)
 \frac {2\,\Ga(5/2-u)} {\pi^{1/2}\,\Ga(1-u)\,(1-r^2)^u}
\,d\vartheta\,dr.
\end{multline}
Using (\ref{e8b}), (\ref{e8d}) and (\ref{e9}) this can be further simplified to
\begin{multline}\label{e10} 
\de_{n_{\in\notin},n_{\notin\in}}\frac {1} {2^n}
\sum _{j,k\ge0} ^{}
 (-1)^{j}
\binom {n_{\in\in}}j \binom
{n_{\notin\notin}-n_{\in\in}}{2k}  \frac {2\,(2j+2k-1)!!\,(2n_{\notin\in})!!}
{(2j+2k+2n_{\notin\in}+1)!!}\\
\cdot\frac
{\Ga(j+k+n_{\notin\in}+3/2)\,\Ga(1-u)}
{2\,\Ga(j+k+n_{\notin\in}+5/2-u)}
 \frac {2\,\Ga(5/2-u)} {\pi^{1/2}\,\Ga(1-u)}.
\end{multline}
Next we interchange sums over $j$ and $k$ 
and write the sum over $k$ in terms of the standard
hypergeometric notation
$${}_r F_s\!\left[\begin{matrix} a_1,\dots,a_r\\ b_1,\dots,b_s\end{matrix}; 
z\right]=\sum _{k=0} ^{\infty}\frac {\po{a_1}{k}\cdots\po{a_r}{k}}
{k!\,\po{b_1}{k}\cdots\po{b_s}{k}} z^k\ ,$$
where the shifted factorial
$(a)_k$ is given by $(a)_k:=a(a+1)\cdots(a+k-1)$,
$k\ge1$, $(a)_0:=1$. Thus we can write (\ref{e10}) in the form
\begin{multline}\label{e11} 
\de_{n_{\in\notin},n_{\notin\in}}\frac {1} {2^n}
\sum _{k\ge0} ^{}\binom {n_{\notin\notin}-n_{\in\in}}{2k}
\frac {(2k-1)!!\,n_{\notin\in}!\,\Ga\(\frac {5} {2}-u\)} {2^{k+1}\,\Ga\(\frac
{5} {2}+k+n_{\notin\in}-u\)}\\
\cdot{}_2F_1\!\[\begin{matrix} \frac {1} {2}+k,-n_{\in\in}\\ \frac {5} 
{2}+k+n_{\notin\in}-u
\end{matrix}; 1\].
\end{multline}
The $_2F_1$ series can be summed by means of Gau\ss' $_2F_1$
summation (see e.g\@. \cite[(1.7.6); Appendix (III.3)]{SlatAC})
\begin{equation}\label{e12}
{} _{2} F _{1} \!\left [ \begin{matrix} { a, b}\\ { c}\end{matrix} ; 
{\displaystyle
   1}\right ]  = \frac {\Ga(c)\,\Ga(c-a-b)} {\Ga(c-a)\,\Ga(c-b)} ,
\end{equation}
provided the series terminates or $\Re (c-a-b)\ge0$. Applying (\ref{e12}) to
the $_2F_1$ in (\ref{e11}) (observe that it is terminating)
and writing the sum over $k$ as a hypergeometric
series, the
expression (\ref{e11}) becomes
\begin{multline}\notag\de_{n_{\in\notin},n_{\notin\in}}\frac {1} {2^n}
\frac {\Ga(2+n_{\in\in}+n_{\notin\in}-u)\,\Ga\(\frac {5}
{2}-u\)\, n_{\notin\in}!} {\Ga\(\frac {5} {2}+n_{\in\in}+n_{\notin\in}-u\)
\,\Ga(2+n_{\notin\in}-u)}\\
\times {}_2F_1\!\[\begin{matrix} \frac {n_{\in\in}} {2}-\frac {n_{\notin\notin}} 
{2},\frac {1} {2}+\frac {n_{\in\in}} {2}-\frac {n_{\notin\notin}} 
{2}\\\frac {5} {2}+n_{\in\in}+n_{\notin\in}-u\end{matrix}; 1\].
\end{multline}
Another application of (\ref{e12}) gives
\begin{multline}\label{e13}
\de_{n_{\in\notin},n_{\notin\in}}\frac {1} {2^n}\\
\times
\frac {\Ga(2+n_{\in\in}+n_{\notin\in}-u)\,
\Ga(2+n_{\notin\notin}+n_{\notin\in}-u)\,\Ga\(\frac {5}
{2}-u\)\, n_{\notin\in}!} {\Ga\(\frac {5} {2}+\frac {n_{\in\in}} {2}+
\frac {n_{\notin\notin}} {2}+n_{\notin\in}-u\)\,\Ga\( {2}+\frac {n_{\in\in}} {2}+
\frac {n_{\notin\notin}} {2}+n_{\notin\in}-u\)
\,\Ga(2+n_{\notin\in}-u)}.\\
\end{multline}
Trivially, we have $n=n_{\in\in}+n_{\notin\notin}+n_{\notin\in}
+n_{\in\notin}$.
Since (\ref{e13}) vanishes unless $n_{\notin\in}=n_{\in\notin}$, we
can substitute
$(n-n_{\in\in}-n_{\notin\notin})/2$ for $n_{\notin\in}$ in the
arguments of the gamma functions. Thus,
we see that (\ref{e13}) equals (\ref{e4}). This completes the proof of the 
Theorem.\quad \quad
\qed
\medskip

\subsection{Eigenvalues and eigenvectors of the Bayesian density
matrices $\ze_n(u)$}\label{s2.2}

With the explicit description of the result $\ze_n(u)$ of averaging 
$\overset n\otimes\rho$ with respect to $q_{u}$
at our disposal, we now proceed to describe the eigenvalues and eigenspaces
of $\ze_n(u)$. The eigenvalues are given in
Theorem~\ref{t2}. Lemma~\ref{l4} gives a complete set of eigenvectors
of $\ze_n(u)$. The reader should note that, though complete,
this is simply a set of linearly independent eigenvectors and not a
fully orthogonal set.
\begin{theorem}\label{t2}
The eigenvalues of the $2^n\times2^n$ matrix 
$\ze_n(u)$, the entries of which are given by {\em(\ref{e4})},
are
\begin{equation}\label{e14}
\la_h=\frac {1} {2^n}\frac {\Ga\(\frac {5} {2}-u\)\, \Ga(2+n-h-u)\,
\Ga(1+h-u)} {\Ga\(\frac {5} {2}+\frac {n} {2}-u\)\, \Ga(2+\frac {n}
{2}-u)\,\Ga(1-u)}, \quad h=0,1,\dots,\fl{\frac {n} {2}},
\end{equation}
with respective multiplicities
\begin{equation}\label{e15}
\frac {(n-2h+1)^2} {(n+1)}\binom {n+1}h.
\end{equation}
\end{theorem}
The Theorem will follow from a sequence of Lemmas. We state the
Lemmas first, then prove Theorem~\ref{t2} assuming the truth of the Lemmas,
and after that provide proofs of the Lemmas.

In the first Lemma some eigenvectors of the matrix $\ze_n(u)$ are
described. Clearly, since $\ze_n(u)$ is a $2^n\times2^n$ matrix, the
eigenvectors are in $2^n$-dimensional space. As we did previously, we
index coordinates by subsets of $[n]$, so that a generic
vector is $(x_S)_{S\in[n]}$. In particular,
given a subset $T$ of $[n]$, the symbol
$e_T$ denotes the standard unit vector with a 1 in the $T$-th
coordinate and 0 elsewhere, i.e.,
$e_T=(\de_{S,T})_{S\in[n]}$. 

Now let $h,s$ be integers with $0\le h\le s \le n-h$ and let $A$ and
$B$ be
two disjoint $h$-element subsets $A$ and $B$ of $[n]$. Then we 
define the vector $v_{h,s}(A,B)$ by
\begin{equation}\label{e16}
v_{h,s}(A,B):=\underset {Y\subseteq [n]\backslash (A\cup B),\ 
\v{Y}=s-h}
{\sum _{X\subseteq A} ^{}}(-1)^{\v{X}}\, e_{X\cup X'\cup
Y},
\end{equation}
where $X'$ is the ``{\it complement of $X$ in $B$}" by which we mean that
if $X$ consists of the $i_1$-, $i_2$-, \dots -largest elements of $A$,
$i_1<i_2<\cdots$, then $X'$ consists of all elements of $B$ {\it
except for} the $i_1$-, $i_2$-, \dots -largest elements of $B$.
For example, let $n=7$. Then the vector $v_{2,3}(\{1,3\},\{2,5\})$ is
given by
\begin{multline}\label{e17}
 e_{\{2,4,5\}}+ e_{\{2,5,6\}}+
e_{\{2,5,7\}}- e_{\{1,4,5\}}- e_{\{1,5,6\}}- e_{\{1,5,7\}}\\
-
e_{\{2,3,4\}}- e_{\{2,3,6\}}- e_{\{2,3,7\}}+ e_{\{1,3,4\}}+
e_{\{1,3,6\}}+ e_{\{1,3,7\}}.
\end{multline}
(In this special case, the possible subsets $X$ of $A=\{1,3\}$ 
in the sum in (\ref{e16}) are $\emptyset$, 
$\{1\}$, $\{3\}$, $\{1,3\}$, with corresponding complements in
$B=\{2,5\}$ being $\{2,5\}$, $\{5\}$, $\{2\}$, $\emptyset$,
respectively,
and the possible sets $Y$ are $\{4\}$, $\{6\}$, $\{7\}$.) Observe
that all sets $X\cup X'\cup Y$ which occur as indices in (\ref{e16}) 
have the same cardinality $s$.

\begin{lemma}\label{l3} Let $h,s$ be integers with $0\le h\le s \le n-h$ and
let $A$ and $B$ be disjoint $h$-element subsets of $[n]$. Then
$v_{h,s}(A,B)$ as defined in {\em (\ref{e16})} is an eigenvector of the matrix
$\ze_n(u)$, the entries of which are given by {\em (\ref{e4})}, for
the eigenvalue $\la_h$, where $\la_h$ is given by {\em (\ref{e14})}.
\end{lemma}
We want to show that the multiplicity of $\la_h$ equals the
expression in (\ref{e15}). Of course, Lemma~\ref{l3} gives many more eigenvectors
for $\la_h$. Therefore, in order to describe a basis for the
corresponding eigenspace, we have to restrict the collection of
vectors in Lemma~\ref{l3}. 

We do this in the following way. Fix $h$, $0\le h\le \fl{n/2}$.
Let $P$ be a lattice path in the
plane integer lattice $\Z^2$, starting in $(0,0)$,
consisting of $n-h$ up-steps $(1,1)$ and
$h$ down-steps $(1,-1)$, which never goes below the $x$-axis.
Figure~1 displays an example with $n=7$ and $h=2$. Clearly, the end
point of $P$ is $(n,n-2h)$. We call a lattice path which starts in
$(0,0)$ and never goes below the $x$-axes a {\it ballot path}. (This
terminology is motivated by its relation to the (two-candidate) {\it ballot
problem\/}, see e.g\@. \cite[Ch.~1, Sec.~1]{MohaAE}. An alternative
term for ballot path which is often used is ``Dyck path'', see
e.g\@. \cite[p.~I-12]{vien}.) We will use the
abbreviation ``b.p.'' for ``ballot path'' in displayed formulas.
\vskip10pt
\vbox{
$$
\Gitter(8,5)(0,0)
\Koordinatenachsen(8,5)(0,0)
\Pfad(0,0),3433433\endPfad
\SPfad(7,0),222\endSPfad
\DickPunkt(0,0)
\DickPunkt(7,3)
\Label\l{\scriptstyle1}(1,1)
\Label\r{\scriptstyle2}(1,1)
\Label\l{\scriptstyle3}(3,1)
\Label\l{\scriptstyle4}(4,2)
\Label\r{\scriptstyle5}(4,2)
\Label\l{\scriptstyle6}(6,2)
\Label\l{\scriptstyle7}(7,3)
\hskip3.5cm
$$
\centerline{\small Ballot paths}
\vskip5pt
\centerline{\small Figure 1}
}
\vskip10pt

Given such a lattice path $P$, label the steps from $1$ to $n$, as
is indicated in Figure~1. Then
define $A_P$ to be set of all labels corresponding to the first $h$
up-steps of $P$ and $B_P$ to 
be set of all labels corresponding to the $h$ down-steps of $P$. In
the example of Figure~1 we have for the choice $h=2$ that 
$A_P=\{1,3\}$ and $B_P=\{2,5\}$.
Thus, to each $h$ and $s$, $0\le h\le s\le n-h$, and $P$ as above we
can associate the vector $v_{h,s}(A_P,B_P)$. In our running example
of Figure~1 the vector
$v_{2,3}(P)$ would hence be $v_{2,3}(\{1,3\},\{2,5\})$, the vector in
(\ref{e17}). To have a more concise form of
notation, we will write $v_{h,s}(P)$ for $v_{h,s}(A_P,B_P)$ from now on.

\begin{lemma}\label{l4}
The set of vectors
\begin{equation}\label{e18}
\{v_{h,s}(P):0\le h\le s\le n-h,\ P\text { a ballot path from $(0,0)$
to $(n,n-2h)$}\}
\end{equation}
is linearly independent.
\end{lemma}

The final Lemma tells us how many such vectors $v_{h,s}(P)$ there
are.
\begin{lemma} \label{l5}
The number of ballot paths from $(0,0)$ to $(n,n-2h)$
is $\frac {n-2h+1} {n+1}\binom {n+1}h$. The total number of all
vectors in the set\/ {\em(\ref{e18})} is $2^n$.
\end{lemma}

Now, let us for a moment assume that Lemmas~\ref{l3}--\ref{l5} are already proved.
Then, Theorem~\ref{t2} follows immediately, as it turns out.
\medskip

{\sc Proof of Theorem~\ref{t2}}. Consider the set of vectors in (\ref{e18}). By
Lemma~\ref{l3} we know that it consists of eigenvectors for the matrix
$\ze_n(u)$. In addition, Lemma~\ref{l4} tells us that this set of vectors is
linearly independent. Furthermore, by Lemma~\ref{l5} the number of vectors in
this set is exactly $2^n$, which is the dimension of the space where
all these vectors are contained. Therefore, they must form a basis of
the space. 

Lemma~\ref{l3} says more precisely that $v_{h,s}(P)$ is an eigenvector for
the eigenvalue $\la_h$. From what we already know, this implies that
for fixed $h$ the set
$$
\{v_{h,s}(P): h\le s\le n-h,\ P\text { a ballot path from $(0,0)$
to $(n,n-2h)$}\}$$
forms a basis for the eigenspace corresponding to $\la_h$. Therefore,
the dimension of the eigenspace corresponding to $\la_h$ equals
the number of possible numbers $s$ times the number of possible
lattice paths $P$. This is exactly
$$(n-2h+1)\frac {(n-2h+1)} {(n+1)}\binom {n+1}h,$$
the number of possible lattice paths $P$ being given by the first
statement of Lemma~\ref{l5}.
This expression equals exactly the expression (\ref{e15}).
Thus, Theorem~\ref{t2} is proved. \quad \quad \qed

\bigskip

Now we turn to the proofs of the Lemmas.
\medskip

{\sc Proof of Lemma~\ref{l3}}. Let $h,s$ and $A,B$ be fixed, satisfying the
restrictions in the statement of the Lemma.
We have to show that
$$\ze_n(u)\cdot v_{h,s}(A,B)=\la_h v_{h,s}(A,B).$$
Restricting our attention to the $I$-th component, we see 
from the
definition (\ref{e16}) of $v_{h,s}(A,B)$ that we need to establish 
\begin{equation}\label{e19}
\underset {Y\subseteq [n]\backslash (A\cup B),\ 
\v{Y}=s-h}
{\sum _{X\subseteq A} ^{}}\kern-1cm Z_{I,X\cup X'\cup Y}\,(-1)^{\v{X}}=
\begin{cases} \la_h (-1)^{\v{U}}&\text {if $I$ is of the form $U\cup U'\cup V$}\\
&\text {for some $U$ and $V$, $U\subseteq A$,}\\
&\text {$V\subseteq [n]\backslash (A\cup B)$,
$\v{V}=s-h$}\\
0&\text {otherwise.}\end{cases}
\end{equation}
We prove (\ref{e19}) by a case by case analysis. The first two cases cover
the case ``otherwise" in (\ref{e19}), the third case treats the first alternative
in (\ref{e19}).

\smallskip
{\it Case 1. The cardinality of $I$ is different from $s$}. As we
observed earlier, the cardinality of any set $X\cup X'\cup Y$ which
occurs as index at the left-hand side of (\ref{e19}) equals $s$. The
cardinality of $I$ however is different from $s$. 
As we observed in the Remark after Theorem~\ref{t1}, this implies that
any coefficient $Z_{I,X\cup X'\cup Y}$ on the left-hand side
vanishes. Thus, (\ref{e19}) is proved in this case.

\smallskip
{\it Case 2. The cardinality of $I$ equals $s$, but
$I$ does not have the form $U\cup U'\cup V$
for any $U$ and $V$, $U\subseteq A$,
$V\subseteq [n]\backslash (A\cup B)$, $\v{V}=s-h$}.
Now the sum on the left-hand side of (\ref{e19}) contains nonzero
contributions. We have to show that they cancel each other. We do
this by grouping summands in pairs, the sum of each pair being 0.

Consider a set $X\cup X'\cup Y$ which occurs as index at the
left-hand side of (\ref{e19}). Let $e$ be minimal such that 
\begin{enumerate}
\item[]either: the $e$-th largest element of $A$ and the
$e$-th largest element of $B$ are both in $I$,
\item[]or: the $e$-th largest element of $A$ and the
$e$-th largest element of $B$ are both not in $I$.
\end{enumerate}
That such an $e$ must exist is guaranteed by our assumptions about
$I$. Now consider $X$ and $X'$. If the $e$-th largest element of $A$
is contained in $X$ then the $e$-th largest element of $B$ is not
contained in $X'$, and vice versa. Define a new set $\bX$ by adding
to $X$ the $e$-th largest element of $A$ if it is not already
contained in $X$, respectively by removing it from $X$ if it is
contained in $X$. Then, it is easily checked that 
$$Z_{I,X\cup X'\cup Y}=Z_{I,\bX\cup \bX'\cup Y}.$$
On the other hand, we have $(-1)^{\v{X}}=-(-1)^{\v\bX}$ since the
cardinalities of $X$ and $\bX$ differ by $\pm1$. Both facts combined
give
$$Z_{I,X\cup X'\cup Y}\,(-1)^{\v{X}}+Z_{I,\bX\cup \bX'\cup Y}\,
(-1)^{\v\bX}=0.$$
Hence, we have found two summands on the left-hand side of
(\ref{e19}) which cancel each other. 

Summarizing,
this construction finds for any $X,Y$ sets $\bX,Y$ such that the
corresponding summands on the left-hand side of (\ref{e19}) cancel each
other. Moreover, this construction applied to $\bX,Y$ gives back
$X,Y$. Hence, what the construction does is exactly what we claimed,
namely it groups the summands into pairs which contribute 0 to the
whole sum. Therefore the sum is 0, which establishes (\ref{e19}) in this case
also.

\smallskip
{\it Case 3. $I$ has the form $U\cup U'\cup V$
for some $U$ and $V$, $U\subseteq A$,
$V\subseteq [n]\backslash (A\cup B)$, $\v{V}=s-h$}.
This assumption implies in particular that the cardinality of $I$ is
$s$. From the Remark after the statement of Theorem~\ref{t1} we know that in
our situation $Z_{I,X\cup X'\cup Y}$ depends only on the number of
common elements in $I$ and $X\cup X'\cup Y$. Thus, the left-hand side
in (\ref{e19}) reduces to
\begin{equation}\label{e20}
\sum _{j,k\ge0} ^{}N(j,k)\,(-1)^{\v{U}+j}\,k!
\frac {1} {2^n}\frac {\Ga\(\frac {5} {2}-u\)\, \Ga(2+n-s-u)\,
\Ga(2+s-u)} {\Ga\(\frac {5} {2}+\frac {n} {2}-u\)\, \Ga\(2+\frac {n}
{2}-u\)\Ga(2+k-u)},
\end{equation}
where $N(j,k)$ is the number of sets $X\cup X'\cup Y$,
for some $X$ and $Y$, $X\subseteq A$,
$Y\subseteq [n]\backslash (A\cup B)$, $\v{Y}=s-h$, which have $s-k$
elements in common with $I$, and which have $h-j$ elements in common
with $I\cap(A\cup B)=U\cup U'$. Clearly, we used expression (\ref{e4}) with
$n_{\in\in}=s-k$ and $n_{\notin\notin}=n-s-k$.

To determine $N(j,k)$, note first that there are $\binom hj$
possible sets $X\cup X'$ which intersect $U\cup U'$ 
in exactly $h-j$ elements. Next, let us assume that we already made a
choice for $X\cup X'$.
In order to determine the number of possible sets $Y$ such that
$X\cup X'\cup Y$ has $s-k$ elements in common with $I$, we have to
choose $(s-k)-(h-j)=s-h+j-k$ elements from
$V$, for which we have $\binom {s-h}{s-h+j-k}$ possibilities, and
we have to choose $s-h-(s-h+j-k)=k-j$ elements from $[n]\backslash
(I\cup A\cup B)$ to obtain a total number of $s$ elements, for which
we have $\binom {n-s-h}{k-j}$ possibilities. Hence,
\begin{equation}\label{e21}
N(j,k)=\binom hj\binom {s-h}{k-j}\binom {n-s-h}{k-j}.
\end{equation}

So it remains to evaluate the double sum (\ref{e20}), using the
expression (\ref{e21})
for $N(j,k)$. 

We start by writing the sum over $j$ in (\ref{e20}) in hypergeometric
notation,
\begin{multline}\notag 
(-1)^{\v{U}}\frac {1} {2^n}
{\frac{\Gamma({ \textstyle {\frac5 2} - u})  \,
     \Gamma({ \textstyle 2 + n - s - u})  \,\Gamma({ \textstyle 2 + s - u})  
}
{\Gamma({ \textstyle 2 - u})  \,
     \Gamma({ \textstyle 2 + {\frac n 2} - u})  \,
     \Gamma({ \textstyle {\frac5 2} + {\frac n 2} - u}) }}\\
\times
{     \sum_{k = 0}^{\infty}
        {\frac{  
            ({ \textstyle h - s}) _{k} \, ({ \textstyle h - n + s}) _{k} }
            {({ \textstyle 1}) _{k} \, ({ \textstyle 2 - u}) _{k} }}
{} _{3} F _{2} \!\left [ \begin{matrix} { -k, -k, -h}\\ { 1 - h - k + s, 1
             - h - k + n - s}\end{matrix} ; {\displaystyle 1}\right ]
}.
\end{multline}
To the $_3F_2$ series we apply a transformation formula of Thomae
(see e.g\@. \cite[(3.1.1)]{GaRaAA}),
\begin{equation}\label{e22}
{} _{3} F _{2} \!\left [ \begin{matrix} { a, b, -m}\\ { d, e}\end{matrix} ;
   {\displaystyle 1}\right ]  =
{\frac{ ({ \textstyle -b + e}) _{m} }
      {({ \textstyle e}) _{m} }}  
{} _{3} F _{2} \!\left [ \begin{matrix} { -m, b, -a + d}\\ { d, 1 + b - e -
       m}\end{matrix} ; {\displaystyle 1}\right ] 
\end{equation}
where $m$ is a nonnegative integer. We write the resulting $_3F_2$
again as a sum over $j$, then interchange sums over $k$ and $j$, and
write the (now) inner sum over $k$ in hypergeometric notation. Thus
we obtain
\begin{multline}\notag 
(-1)^{\v{U}}\frac {1} {2^n}
{\frac{\Gamma({ \textstyle {\frac5 2} - u})  \,
     \Gamma({ \textstyle 2 + n - s - u})  \,\Gamma({ \textstyle 2 + s - u})  
}
{    \Gamma({ \textstyle {\frac5 2} + {\frac n 2} - u}) \,
     \Gamma({ \textstyle 2 + {\frac n 2} - u})  \,
   \Gamma({ \textstyle 2 - u})}}\\
\times{ \sum_{j = 0}^{\infty}
        {\frac{  
            ({ \textstyle -h}) _{j} \, ({ \textstyle 1 - h + s}) _{j} } 
          {({ \textstyle 1}) _{j} \, ({ \textstyle 2 - u}) _{j} }}  }
{} _{2} F _{1} \!\left [ \begin{matrix} { j - n + s, h - s}\\ { 2 + j -
             u}\end{matrix} ; {\displaystyle 1}\right ].
\end{multline}
The $_2F_1$ series in this expression is terminating because $h-s$ is
a nonpositive integer. Hence, it can be summed by means of Gau\ss' sum
(\ref{e12}).
Writing the remaining sum over $j$ in hypergeometric notation, the
above expression becomes
$$(-1)^{\v{U}}\frac {1} {2^n}
{\frac{  
     \Gamma({ \textstyle {\frac5 2} - u})  \,
     \Gamma({ \textstyle 2 + n - h - u}) \, \Gamma({ \textstyle 2 + s - u}) }
     {      \Gamma({ \textstyle {\frac5 2} + {\frac n 2} - u})  \,
\Gamma({ \textstyle 2 + {\frac n 2} - u})  \,
     \Gamma({ \textstyle 2 + s -h - u}) }}
{} _{2} F _{1} \!\left [ \begin{matrix} { -h, 1 - h + s}\\ { 2 - h + s -
      u}\end{matrix} ; {\displaystyle 1}\right ].
$$
Again, 
the $_2F_1$ series is terminating and so
is summable by means of (\ref{e12}). Thus, we get
$$(-1)^{\v{U}}\frac {1} {2^n}\frac {\Ga\(\frac {5} {2}-u\)\, \Ga(2+n-h-u)\,
\Ga(1+h-u)} {\Ga\(\frac {5} {2}+\frac {n} {2}-u\)\, \Ga(2+\frac {n}
{2}-u)\,\Ga(1-u)},$$
which is exactly the expression (\ref{e14}) for $\la_h$ times $(-1)^{\v{U}}$. 
This proves (\ref{e19}) in this case.

\smallskip
The proof of Lemma~\ref{l3} is now complete.\quad \quad \qed
\medskip

\medskip
{\sc Proof of Lemma~\ref{l4}}.
We know from Lemma~\ref{l3} that $v_{h,s}(P)$ lies in the eigenspace for the
eigenvalue $\la_h$, with $\la_h$ being given in (\ref{e14}). The $\la_h$'s,
$h=0,1,\dots, \fl{n/2}$, are all distinct, so the corresponding
eigenspaces are linearly independent. Therefore it suffices to show
that for any {\it fixed\/} $h$ the set of vectors
$$\{v_{h,s}(P): h\le s\le n-h,\ P\text { a ballot path from $(0,0)$
to $(n,n-2h)$}\}$$
is linearly independent.

On the other hand, a vector $v_{h,s}(A,B)$ lies in the space spanned
by the standard unit vectors $e_T$ with $\v{T}=s$. Clearly, as $s$
varies, these spaces are linearly independent. Therefore, it suffices to
show that for any {\it fixed\/} $h$ {\it and\/} $s$ the set of vectors
$$\{v_{h,s}(P): P\text { a ballot path from $(0,0)$
to $(n,n-2h)$}\}$$
is linearly independent.

So, let us fix integers $h$ and $s$ with $0\le h\le s\le n-h$, and
let us suppose that there is some vanishing linear combination
\begin{equation}\label{e23}
\sum _{P\text{ b.p\@. from }(0,0)\text{ to }(n,n-2h)} ^{}
c_P\,v_{h,s}(P)=0.
\end{equation}
We have to establish that $c_P=0$ for all ballot paths $P$ from
$(0,0)$ to $(n,n-2h)$. 

We prove this fact by induction on the set of ballot paths from
$(0,0)$ to $(n,n-2h)$. In order to make this more precise, we need to
impose a certain order on the ballot paths. Given a ballot path $P$
from $(0,0)$ to $(n,n-2h)$, we define its {\it front portion\/} $F_P$
to be the portion of $P$ from the beginning up to and including $P$'s
$h$-th up-step. For example, choosing $h=2$, the front portion of the
ballot path in Figure~1 is the subpath from $(0,0)$ to $(3,1)$. Note
that  $F_P$ can be any ballot path starting in $(0,0)$ with
$h$ up-steps and less than $h$ down-steps. We order such front
portions lexicographically, in the sense that $F_1$ is before $F_2$ if
and only if $F_1$ and $F_2$ agree up to some point and then $F_1$
continues with an up-step while $F_2$ continues with a down-step.

Now, here is what we are going to prove: Fix any possible front
portion $F$. We shall show that $c_P=0$ for all $P$ with front portion
$F_P$ equal to $F$, {\it given that it is already known that
$c_{P'}=0$ for all $P'$ with a front portion $F_{P'}$ that is before
$F$.} Clearly, by induction, this would prove $c_P=0$ for {\it all\/}
ballot paths $P$ from $(0,0)$ to $(n,n-2h)$.

Let $F$ be a possible front portion, i.e., a ballot path starting in
$(0,0)$ with exactly $h$ up-steps and less than $h$ down-steps. As we
did earlier, label the steps of $F$ by $1,2,\dots$, and denote the set
of labels corresponding to the down-steps of $F$ by $B_F$. We write
$b$ for $\v{B_F}$, the number of all down-steps of $F$. Observe that
then the total number of steps of $F$ is $h+b$. 

Now, let $T$ be a fixed $(h-b)$-element subset of
$\{h+b+1,h+b+2,\dots,n\}$. Furthermore, let $S$ be a set of the form
$S=B_F\cup S_1\cup S_2$, where $S_1\subseteq T$ and $S_2\subseteq
\{h+b+1,h+b+2,\dots,n\}\backslash T$, and such that $\v{S}=s$. 

We consider the coefficient of $e_S$ in the left-hand side of (\ref{e23}). To
determine this coefficient, we have to determine the coefficient of
$e_S$ in $v_{h,s}(P)$, for all $P$. We may concentrate on those $P$
whose front portion $F_P$ is equal to or later than $F$, since our
induction hypothesis says that $c_P=0$ for all $P$ with $F_P$ before
$F$. So, let $P$ be a ballot path from $(0,0)$ to $(n,n-2h)$ with
front portion equal to or later than $F$. We claim that the
coefficient of $e_S$ in $v_{h,s}(P)$ is zero unless the set $B_P$ of
down-steps of $P$ is contained in $S$.

Let the coefficient of $e_S$ in $v_{h,s}(P)$ be nonzero.
To establish the claim, we first prove that the front portion $F_P$
of $P$ has to equal $F$. Suppose that this is not the case. Then the
front portion of $P$ runs in parallel with $F$ for some time, say for
the first $(m-1)$ steps, with some $m\le h+b$, and then $F$ continues
with an up-step and $F_P$ continues with a down-step (recall that
$F_P$ is equal to or later than $F$). By (\ref{e16}) we have
\begin{equation}\label{e24}
v_{h,s}(P):=\underset {Y\subseteq [n]\backslash (A_P\cup B_P),\ 
\v{Y}=s-h}
{\sum _{X\subseteq A_P} ^{}}(-1)^{\v{X}}\, e_{X\cup X'\cup
Y}.
\end{equation}
We are assuming that the coefficient of $e_S$ in $v_{h,s}(P)$ is
nonzero, therefore $S$ must be of the form $S=X\cup X'\cup Y$, with
$X,Y$ as described in (\ref{e24}). We are considering the case that the
$m$-th step of $F_P$ is a down-step, whence $m\in B_P$, while the
$m$-th step of $F$ is an up-step, whence $m\notin B_F$. By definition
of $S$, we have $S\cap \{1,2\dots,h+b\}=B_F$, whence $m\notin S$. 

Summarizing so far, we have $m\in B_P$, $m\notin S$, for some $m\le
h+b$, and $S=X\cup X'\cup Y$, for some $X,Y$ as described in (\ref{e24}). In
particular we have $m\notin X'$. Now recall that $X'$ is the
``complement of $X$ in $B_P$''. This says in particular that, if $m$ is
the $i$-th largest element in $B_P$, then the $i$-th largest element
of $A_P$, $a$ say, is an element of $X$, and so of $S$. By
construction of $A_P$ and $B_P$, $a$ is smaller than $m$, so in
particular $a<h+b$. As we already observed, there holds $S\cap
\{1,2,\dots,h+b\}=B_F$, so we have $a\in B_F$, i.e., the $a$-th step
of $F$ is a down-step. On the other hand, we assumed that $P$ and $F$
run in parallel for the first $(m-1)$ steps. Since $a\in A_P$, the set
of up-steps of $P$, the $a$-th step of $P$ is an up-step. We have
$a\le m-1$, therefore the $a$-th step of $F$ must be an up-step
also. This is absurd. Therefore, given that the
coefficient of $e_S$ in $v_{h,s}(P)$ is nonzero, the front portion
$F_P$ of $P$ has to equal $F$.

Now, let $P$ be a ballot path from $(0,0)$ to $(n,n-2h)$ with
front portion equal to $F$, and suppose that $S$ has the form $S=X\cup
X'\cup Y$, for some $X,Y$ as described in (\ref{e24}). By definition of the
front portion, the set $A_P$ of up-steps of $P$ has the property
$A_P\cap \{1,2,\dots,h+b\}=\{1,2,\dots,h+b\}\backslash B_F$. Since
$\v{B_F}=b$, these are the labels of exactly $h$ up-steps. Since the
cardinality of $A_P$ is exactly $h$ by definition, we must have
$A_P=\{1,2,\dots,h+b\}\backslash B_F$. Because of $S\cap
\{1,2,\dots,h+b\}=B_F$, which we already used a number of times, $A_P$
and $S$ are disjoint, which in particular implies that $A_P$ and $X$
are disjoint. However, $X$ is a subset of $A_P$ by definition, so $X$
must be empty. This in turn implies that $X'=B_P$. This says nothing
else but that the set $B_P$ of down-steps of $P$ equals $X'$ and so is
contained in $S$. This establishes our claim. 

In fact, we proved more. We saw that $S$ has the form $S=X\cup X'\cup
Y$, with $X=\emptyset$. This implies that the coefficient of $e_S$ in
$v_{h,s}(P)$, as given by (\ref{e24}), is actually $+1$. Comparison of
coefficients of $e_S$ in (\ref{e23}) then gives
\begin{equation}\label{e25}
\underset {F_P=F,\ B_P\subseteq S}
{\sum _{P\text{ b.p\@. from }(0,0)\text{ to }(n,n-2h)} ^{}}
c_P=0,
\end{equation}
for any $S=B_F\cup S_1\cup S_2$, where $S_1\subseteq T$ and $S_2\subseteq
\{h+b+1,h+b+2,\dots,n\}\backslash T$, and such that $\v{S}=s$.

Now, we sum both sides of (\ref{e25}) over all such sets $S$, keeping the
cardinality of $S_1$ and $S_2$ fixed, say
$\v{S_1}=h-b-j$, enforcing $\v{S_2}=s-h+j$, for a fixed $j$, $0\le
j\le h-b$. For a fixed ballot path $P$ from $(0,0)$ to $(n,n-2h)$,
with front portion $F$, with $h-b-k$ down-steps in $T$, and hence with
$k$ down-steps in $\{h+b+1,h+b+2,\dots,n\}\backslash T$, there are
$\binom k{k-j}$ such sets $S_1\subseteq T$ containing all the
$h-b-k$ down-steps of $P$ in $T$, and there are $\binom
{n-(h+b)-(h-b)-k} {s-h+j-k}$ such sets $S_2\subseteq
\{h+b+1,h+b+2,\dots,n\}\backslash T$ containing all the $k$ down-steps
of $P$ in $\{h+b+1,h+b+2,\dots,n\}\backslash T$. Therefore,
summing up (\ref{e25}) gives
\begin{equation}\label{e26}
\sum _{k\ge0} ^{}\binom kj\binom {n-2h-k}{n-h-s-j} 
\bigg(\kern-5pt\underset {\v{B_P\cap(\{h+b+1,h+b+2,\dots,n\}\backslash T)}=k}
{\underset {F_P=F,\ \v{B_P\cap T}=h-b-k}
{\sum _{P\text{ b.p\@. from }(0,0)\text{ to }(n,n-2h)} ^{}}}\kern-.3cm
c_P\bigg)=0,\quad 
j=0,1,\dots,h-b.
\end{equation}
Denoting the inner sum in (\ref{e26}) by $C(k)$, we see that
(\ref{e26}) represents
a non-degenerate triangular system of linear equations for $C(0),
C(1),\dots, C(h-b)$. Therefore, all the quantities $C(0),
C(1),\dots, C(h-b)$ have to equal 0. In particular, we have
$C(0)=0$. Now, $C(0)$ consists of just a single term $c_P$, with $P$
being the ballot path from $(0,0)$ to $(n,n-2h)$, with front portion
$F$, and the labels of the $h-b$ down-steps besides those of $F$
being exactly the elements of $T$. Therefore, we have $c_P=0$ for this ballot
path. The set $T$ was an arbitrary $(h-b)$-subset of
$\{h+b+1,h+b+2,\dots,n\}$. Thus, we have proved $c_P=0$ for any ballot
path $P$ from $(0,0)$ to $(n,n-2h)$ with front portion $F$. This
completes our induction proof. \quad \quad \qed

\medskip

\medskip
{\sc Proof of Lemma~\ref{l5}}. That the number of ballot paths from $(0,0)$
to $(n,n-2h)$ equals $\frac {n-2h+1} {n+1}\binom {n+1}h$ is a
classical combinatorial result (see e.g\@. \cite[Theorem~1
with $t=1$]{MohaAE}). From
this it follows that the total number of vectors in the set (\ref{e18}) is
\begin{equation}\label{e27}
\sum _{h=0} ^{\fl{n/2}}(n-2h+1)\frac {(n-2h+1)} {(n+1)}\binom
{n+1}h.
\end{equation}
To evaluate this sum, note that the summand is invariant under the
substitution $h\to n-2h+1$. Therefore, extending the range of
summation in (\ref{e27}) to $h=0,1,\dots,n+1$ and dividing the result by $2$
gives the same value. So, the cardinality of the set (\ref{e18}) is also
given by
$$\frac {1} {2}\sum _{h=0} ^{n+1}\frac {(n-2h+1)^2} {(n+1)}\binom
{n+1}h.$$
The reader will not have any difficulty in splitting this sum into three
parts so that each part can be summed by means of the binomial
theorem. (Computer algebra systems like {\sl Maple} or
{\sl Mathematica} do this automatically.) The result is exactly
$2^n$, as was claimed.\quad \quad \qed

\medskip

In fact, Theorem~\ref{t2} can be generalized to a wider class of matrices. 

\begin{theorem}\label{t6}
Let $\tilde \ze_n(u)=(\tilde Z_{IJ})_{I,J\in [n]}$ be the $2^n\times
2^n$ matrix defined by
$$\tilde Z_{IJ}:=\de_{n_{\notin\in},n_{\in\notin}} \frac {\big(\tfrac
{n-n_{\in\in}- n_{\notin\notin}} 2\big)! } {\Ga\(2+
\tfrac
{n-n_{\in\in}- n_{\notin\notin}} 2-u\)}\cdot
f(n_{\in\in}-n_{\notin\notin}),$$
where $n_{\in\in}$, etc., have the same meaning as earlier, and where
$f(x)$ is a function of $x$ which is symmetric, i.e., $f(x)=f(-x)$.
Then, the eigenvalues of 
$\tilde\ze_n(u)$ are
\begin{equation}\label{e28}
\la_{h,s}=f(n-2s)\frac { \Ga(2+n-h-u)\,
\Ga(1+h-u)} {\Ga(2+n-s-u) \,\Ga(2+s-u)\,
\Ga(1-u)}, \quad 0\le h\le s\le n-h,
\end{equation}
with respective multiplicities
\begin{equation}\label{e29}
\frac {n-2h+1} {n+1}\binom {n+1}h,
\end{equation}
independent of $s$.

\end{theorem}
\medskip

{\sc Proof}.
The above proof of Theorem~\ref{t2} has to be adjusted only
insignificantly to yield a proof of Theorem~\ref{t6}. In particular, the
vector $v_{h,s}(A,B)$ as defined in (\ref{e16}) is an eigenvector for
$\la_{h,s}$, for any two disjoint $h$-element subsets $A$ and $B$ 
of $[n]$, and
the set (\ref{e18}) is a basis of eigenvectors for $\tilde\ze_n(u)$.
\quad \quad \qed
\medskip

\subsection{The relative entropies of $
 \raise6pt\hbox{$^n$}\kern-9pt \otimes\rho$ 
with
respect to the Bayesian density matrices $\ze_n(u)$}\label{s2.3}

We now apply the preceding results to compute the relative entropy 
$S(\overset n\otimes \rho,\ze_n(u))$ of
$\overset n\otimes \rho$ with respect to $\ze_n(u)$. Utilizing the
definition (\ref{eq:5}) of relative entropy and employing the property
\cite{oh,wehrl} that $S(\overset n\otimes\rho)=nS(\rho)$,
it is given by
\begin{equation}\label{e31}
-n\,S(\rho)-\Tr\(\overset
n\otimes \rho\cdot \log\ze_n(u)\).
\end{equation}
For the first term, for the entropy $S(\rho)$ of $\rho$, $\rho$ being
given by (\ref{eq:6}), 
we have, using spherical coordinates $(r,\vartheta,\phi)$, so that 
$r=(x^2+y^2+z^2)^{1/2}$,
\begin{equation} \label{eq:7}
S(\rho)=-{\frac{(1-r)}  2}{ \log {\frac{(1-r)}  2}} -{\frac {(1+r) }
2}{ \log {\frac{(1+r)}   2}} .
\end{equation}
Concerning the
second term in (\ref{e31}), we have the following theorem.
\begin{theorem}\label{t7}
Let $\ze_n(u)=(Z_{IJ})_{I,J\in[n]}$ be the matrix with
entries $Z_{IJ}$ given in {\em(\ref{e4})}. Then, we have
\begin{multline}\label{e30}
\Tr\(\overset n\otimes \rho\cdot\log \ze_n(u)\)\\
=\sum _{h=0}
^{\fl{n/2}} \frac {n-2h+1} {n+1}\binom {n+1}h\frac {1} {2^{n+1}r}
\big((1+r)^{n+1-h}(1-r)^h-(1+r)^h(1-r)^{n+1-h}\big)\log\la_h,
\end{multline} 
with $\la_h$ as given in {\em(\ref{e14})},
and with $r=\sqrt{x^2+y^2+z^2}$. 
\end{theorem}
Before we move on to the proof, we note that Theorem~\ref{t7} gives us 
the following expression for the relative entropy of 
$\overset n\otimes \rho$ with respect to $\ze_n(u)$

\begin{corollary}\label{c8}
The relative entropy 
$S(\overset n\otimes \rho,\ze_n(u))$ of 
$\overset n\otimes \rho$ with respect to $\ze_n(u)$ equals
\begin{multline}\label{e32}   
\frac {n} {2}(1-r)\log((1-r)/2)+\frac {n}
{2}(1+r)\log((1+r)/2)\\
-\sum _{h=0} ^{\fl{n/2}}\frac {(n-2h+1)} {(n+1)}\binom {n+1}h\hskip5cm\\
\cdot\frac
{1} {2^{n+1}r}\((1+r)^{n-h+1}(1-r)^h-(1+r)^h(1-r)^{n-h+1}\)\log\la_h,
\end{multline}
with $\la_h$ as given in {\em(\ref{e14})}, 
and with $r=\sqrt{x^2+y^2+z^2}$.  
\end{corollary}

\medskip
{\sc Proof of Theorem~\ref{t7}}.
One way of determining the trace of a linear
operator $L$ is 
to choose a basis of the vector space, $\{v_I:I\in [n]\}$ say,
write the action of $L$ on the basis elements in the form
$$Lv_I=c_Iv_I+\text{linear combination of $v_J$'s, $J\ne I$},$$
and then form the sum $\sum _{I} ^{}c_I$ of the ``diagonal'' 
coefficients, which
gives exactly the trace of $L$.

Clearly, we choose as a basis our set (\ref{e18}) of eigenvectors for
$\ze_n(u)$. To determine the action of $\overset n\otimes\rho
\cdot\log\ze_n(u)$ we need only to find the action of $\overset
n\otimes\rho$ on the vectors in the set (\ref{e18}). We claim that this
action can be described as
\begin{multline}\label{e33}
\big(\overset n\otimes\rho\big)\cdot v_{h,s}(P)\\
=\frac {1} {2^n}\bigg(\sum _{k\ge j\ge0} ^{}
(-1)^j\binom hj\binom {s-h}{k-j}\binom {n-s-h}{k-j}(1+z)^{s-k}
(x^2+y^2)^k(1-z)^{n-s-k}\bigg)\\
\cdot v_{h,s}(P)\ {}+{}\ \text{linear combination of eigenvectors}\\
\text{$v_{h',s'}(P')$ with $s'\ne s$},
\end{multline}
for any basis vector $v_{h,s}(P)$ in (\ref{e18}).

To see this, consider the $I$-th component of $\big(\overset n\otimes
\rho\big)\cdot v_{h,s}(P)$, i.e., the coefficient of $e_I$ in
$\big(\overset n\otimes
\rho\big)\cdot v_{h,s}(P)$, $I\in[n]$. By the definition (\ref{e16}) of
$v_{h,s}(P)$ it equals
\begin{equation}\label{e34}
\underset {Y\subseteq [n]\backslash (A_P\cup B_P),\ 
\v{Y}=s-h}
{\sum _{X\subseteq A_P} ^{}}\kern-1cm R_{I,X\cup X'\cup
Y}\,(-1)^{\v{X}},
\end{equation}
where $R_{IJ}$ denotes the $(I,J)$-entry of $\overset
n\otimes\rho$. (Recall that $R_{IJ}$ is given explicitly in (\ref{e2}).)
Now, it should be observed that we did a similar calculation already,
namely in the proof of Lemma~\ref{l3}. In fact, the expression (\ref{e34}) is 
almost
identical with the left-hand side of (\ref{e19}). The essential difference is
that $Z_{IJ}$ is replaced by $R_{IJ}$ for all $J$ (the nonessential
difference is that $A,B$ are replaced by $A_P,B_P$, respectively).  
Therefore, we can partially rely upon
what was done in the proof of Lemma~\ref{l3}. 

We distinguish between the same cases as in the proof of Lemma~\ref{l3}.

\smallskip
{\it Case 1. The cardinality of $I$ is different from $s$}. We do not
have to worry about this case, since $e_I$ then lies in the span of
vectors $v_{h',s'}(P')$ with $s'\ne s$, which is taken care of in
(\ref{e33}). 

\smallskip
{\it Case 2. The cardinality of $I$ equals $s$, but
$I$ does not have the form $U\cup U'\cup V$
for any $U$ and $V$, $U\subseteq A_P$,
$V\subseteq [n]\backslash (A_P\cup B_P)$, $\v{V}=s-h$}.
Essentially the same arguments as those in Case~2 in the proof of
Lemma~\ref{l3} show that the term (\ref{e34}) vanishes for this choice of $I$. Of
course, one has to use the explicit expression (\ref{e2}) for $R_{IJ}$.

\smallskip
{\it Case 3. $I$ has the form $U\cup U'\cup V$
for some $U$ and $V$, $U\subseteq A_P$,
$V\subseteq [n]\backslash (A_P\cup B_P)$, $\v{V}=s-h$}.
In Case~3 in the proof of Lemma~\ref{l3} we observed that there are
 $N(j,k)$ sets $X\cup X'\cup Y$,
for some $X$ and $Y$, $X\subseteq A_P$,
$Y\subseteq [n]\backslash (A_P\cup B_P)$, $\v{Y}=s-h$, which have $s-k$
elements in common with $I$, and which have $h-j$ elements in common
with $I\cap(A_P\cup B_P)=U\cup U'$, where $N(j,k)$ is given by
(\ref{e21}). Then, using the explicit expression (\ref{e2}) for $R_{IJ}$,
it is straightforward to see that the expression (\ref{e34}) equals
$$\frac {1} {2^n}\sum _{k\ge j\ge0} ^{}
(-1)^{\v U+j}\binom hj\binom {s-h}{k-j}\binom {n-s-h}{k-j}(1+z)^{s-k}
(x^2+y^2)^k(1-z)^{n-s-k}$$
in this case. This establishes (\ref{e33}).

\smallskip
Now we are in the position to write down an expression for the trace
of $\overset n\otimes\rho\cdot\log\ze_n(u)$. By Theorem~\ref{t2} and by 
(\ref{e33})
we have
\begin{multline}\label{e35}
\(\overset n\otimes\rho\cdot \log\ze_n(u)\)
\cdot v_{h,s}(P)\\
=\frac {1} {2^n}\bigg(\sum _{k\ge j\ge0} ^{}
(-1)^j\binom hj\binom {s-h}{k-j}\binom {n-s-h}{k-j}(1+z)^{s-k}
(x^2+y^2)^k(1-z)^{n-s-k}\bigg)\\
\cdot \log\la_h\cdot v_{h,s}(P)+\text{linear combination of eigenvectors}\\
\text{$v_{h',s'}(P')$ with $s'\ne s$}.
\end{multline}
 From what was said at the beginning of this proof, in order
to obtain the trace of $\overset n\otimes\rho\cdot\log\ze_n(u)$, we have to
form the sum of all the ``diagonal'' coefficients in (\ref{e35}).
Using the first statement of Lemma~\ref{l5} and replacing $x^2+y^2$ by
$r^2-z^2$, we see that it is
\begin{multline}\label{e36}
\sum _{h=0} ^{\fl{n/2}}\log\la_h \,\frac {(n-2h+1)} {(n+1)}\binom
{n+1}h\frac {1} {2^n}
\sum _{s=h} ^{n-h}\sum _{k\ge j\ge0} ^{}
(-1)^j\binom hj\binom {s-h}{k-j}\binom {n-s-h}{k-j}\\
\cdot (1+z)^{s-k}
(r^2-z^2)^k(1-z)^{n-s-k}.
\end{multline}
In order to see that this expression equals (\ref{e30}), we have to prove
\begin{multline}\label{e37} 
\sum _{s=h} ^{n-h}\sum _{j=0} ^{h}\sum _{k=j} ^{s}
(-1)^j\binom hj\binom {s-h}{k-j}\binom {n-s-h}{k-j} (1+z)^{s-k}
(r^2-z^2)^k(1-z)^{n-s-k}\\
=\frac {1} {2r}
\big((1+r)^{n+1-h}(1-r)^h-(1+r)^h(1-r)^{n+1-h}\big).
\end{multline}

We start with the left-hand side of (\ref{e37}) and write the inner sum in
hypergeometric notation, thus obtaining
$$\sum_{s = h}^{ n-h}\sum_{j = 0}^{h}
     {{\left( 1 - z \right) }^{ n - s-j}} 
         {{\left( 1 + z \right) }^{s-j}} 
         {{\left( {r^2} - {z^2} \right) }^j} 
        {\frac{     ({ \textstyle -h}) _{j} } {({ \textstyle 1}) _{j} }}
     {} _{2} F _{1} \!\left [ \begin{matrix} {  h - n + s,h-s}\\ { 1}\end{matrix}
        ; \frac {r^2-z^2} {1-z^2}\right].
$$
To the $_2F_1$ series we apply the transformation formula
(\cite[(1.8.10),
terminating form]{SlatAC}
$$
{} _{2} F _{1} \!\left [ \begin{matrix} { a, -m}\\ { c}\end{matrix} ; 
{\displaystyle
   z}\right ]  = 
{\frac{    ({ \textstyle c-a}) _{m} } {({ \textstyle c}) _{m} }}
{} _{2} F _{1} \!\left [ \begin{matrix} { -m, a}\\ { 1 + a - c -
       m}\end{matrix} ; {\displaystyle 1 - z}\right ] , 
$$
where $m$ is a nonnegative integer. We write the resulting $_2F_1$
series again as a sum over $k$. In the resulting expression
we exchange sums so that the sum over $j$ becomes the innermost sum.
Thus, we obtain
\begin{multline}\notag
 \sum_{s = h}^{n-h}\sum_{k = 0}^{s-h}
     {{\left( 1 - r^2 \right) }^k}
         {{\left( 1 - z \right) }^{n - s-k}} 
         {{\left( 1 + z \right) }^{s-k}} \\
\cdot       {\frac{  ({ \textstyle h - s}) _{k}\,  
         ({ \textstyle n - h - s+1}) _{s-h}\,  
         ({ \textstyle h - n + s}) _{k} } 
       { ({ \textstyle 1}) _{k}\,  
         ({ \textstyle 1}) _{s-h}  \,({ \textstyle 2 h - n}) _{k} }}
\sum _{j=0} ^{h}\binom hj \(\frac {z^2-r^2} {1-z^2}\)^j.
\end{multline}
Clearly, the innermost sum can be evaluated by the binomial theorem.
Then, we interchange sums over $s$ and $k$. The expression
that results is
\begin{multline}\notag \sum_{k = 0}^{\fl{n/2}-h }
{{\left( 1 - r^2 \right) }^{h + k}} 
       {{\left( 1 - z \right) }^{n-2h - 2k }} 
{\frac{       ({ \textstyle 2 h + k - n}) _{k} } {({ \textstyle 1}) _{k} }}\\
\cdot \sum _{s=0} ^{n-2h-2k}\binom {n-2h-2k}s \(\frac {1+z} {1-z}\)^s.
\end{multline}
Again, we can apply the binomial theorem.
Thus,  we reduce our expression on the left-hand
side of (\ref{e37}) to
$${2^{n-2 h}} {{\left( 1 - r^2 \right) }^h} 
 \sum _{k=0} ^{\fl{n/2}-h}\frac {\(h-\frac {n} {2}\)_k\, \(h-\frac
{n} {2}+\frac {1} {2}\)_k} {(2h-n)_k\, k!}(1-r^2)^k.$$
Now, we replace $(1-r^2)^k$ by its binomial expansion $\sum _{l=0}^k(-1)^l
\binom kl r^{2l}$, interchange sums over $k$ and $l$, and write
the (now) inner sum over $k$ in hypergeometric notation. This gives
\begin{multline}\notag {2^{n-2 h }} {{\left( 1 - r^2 \right) }^h} 
  \bigg( \sum_{l = 0}^{\fl{n/2}-h}{{\left( -1 \right) }^l} {r^{2 l}} 
{\frac{
    ({ \textstyle h - {\frac n 2}}) _{l}\,  
         ({ \textstyle {\frac1 2} + h - {\frac n 2}}) _{l} } 
       {({ \textstyle 1}) _{l}\,  ({ \textstyle 2 h - n}) _{l} }}  \\
     \cdot {} _{2} F _{1} \!\left [ \begin{matrix} { h + l - {\frac n
  2}, {\frac 1 2} + h
          + l - {\frac n 2}}\\ { 2 h + l - n}\end{matrix} ; {\displaystyle
          1}\right ]\bigg) .
\end{multline}
Finally, this $_2F_1$ series can be summed by means of Gau\ss'
summation (\ref{e12}). Simplifying, we have
$$ {{\left( 1 - r ^2\right) }^h}
  \sum _{l=0} ^{\fl{n/2}-h}\binom {n-2h+1}{2l+1}r^{2l},$$
which is easily seen to equal the right-hand side in (\ref{e37}). This
completes the proof of the Theorem. \quad \quad \qed

\medskip

\subsection{Asymptotics of the relative entropy
 of $\raise6pt\hbox{$^n$}\kern-9pt \otimes\rho$ with respect to 
$\ze_n(u)$}\label{s2.4}
In the preceding subsection, we obtained in Corollary~\ref{c8}
the general formula (\ref{e32}) for the relative entropy of $\overset
n\otimes\rho$ with respect to the Bayesian density matrix
$\ze_n(u)$. We, now, proceed to find its asymptotics for $n\to\infty$.
We prove the following theorem.

\begin{theorem}\label{t9}
The asymptotics of the relative entropy 
$S(\overset n\otimes \rho,\ze_n(u))$ of 
$\overset n\otimes \rho$ with respect to $\ze_n(u)$ for a fixed 
$r=\sqrt{x^2+y^2+z^2}$ with $0\le r<1$ is given by
\begin{multline}\label{a3} 
\frac {3} {2}\log n  -\frac {1} {2}- \frac {3} {2} \log 2  
-(1-u)\log(1-r^2)+\frac {1} {2r}\log\(\frac {1-r} {1+r}\)\\
+ \log\Ga(1 - u)  -\log\Ga(5/2 - u)
 +O\(\frac {1} {n}\).
\end{multline}
In the case $r=0$, this means that the asymptotics is given by the
expression {\em(\ref{a3})} in the limit $r\downarrow0$, i.e., by
\begin{equation}\label{a4} 
\frac {3} {2}\log n  -\frac {3} {2}- \frac {3} {2} \log 2  
+ \log\Ga(1 - u) -\log\Ga(5/2 - u) 
 +O\(\frac {1} {n}\).
\end{equation}
For any fixed $\ep>0$, the $O(.)$ term in {\em(\ref{a3})} 
is uniform in $u$ and $r$ as long as $0\le r\le 1-\ep$.

For $r=1$ the asymptotics is given by
\begin{equation}\label{a5}
 (2-u)\log n+(2u-3)\log 2
+ \frac {1} {2} \log\pi - \log\Ga(5/2 - u) + O\(\frac {1} {n}\).
\end{equation}
Also here, the $O(.)$ term is uniform in $u$.
\end{theorem}
\begin{remark}\em 
It is instructive to observe that, although a comparison of (\ref{a3})
and (\ref{a5}) seems to suggest that the
asymptotics of the relative entropy of 
$\overset n\otimes \rho$ with respect to $\ze_n(u)$ behaves completely 
differently for $0\le r<1$ and $r=1$, the two cases are really quite
compatible. In fact, letting $r$ tend to $1$ in (\ref{a3}) shows that
(ignoring the error term) the asymptotic expression approaches
$+\infty$ for $u<1/2$, $-\infty$ for $u>1/2$, and it approaches
$ \frac {3} {2}\log n-\frac {1} {2}-\frac {5} {2}\log 2
+ \frac {1} {2} \log\pi $ for $u=1/2$. This indicates that, for
$r=1$, the order of magnitude of the relative entropy of 
$\overset n\otimes \rho$ with respect to $\ze_n(u)$ should be larger
than $\frac {3} {2}\log n$ if $u<1/2$, smaller than $\frac {3}
{2}\log n$ if $u>1/2$, and exactly $\frac {3} {2}\log n$ if $u=1/2$.
How much larger or smaller is precisely what formula (\ref{a5})
tells us: the order of magnitude is $(2-u)\log n$, and in the case $u=1/2$
the asymptotics is, in fact, $ \frac {3} {2}\log n-2\log 2
+ \frac {1} {2} \log\pi $.
\end{remark}
\medskip

{\sc Sketch of Proof of Theorem~\ref{t9}}. 
We have to estimate the expression (\ref{e32}) for large $n$.
Clearly, it suffices to concentrate on the sum in (\ref{e32}). 
Because of $\la_{n+1-h}=\la_h$, this sum can be also expressed as
\begin{equation}\label{a6}
\frac {1} {2^{n+1}r}\sum _{h=0} ^{n+1}\frac {(n-2h+1)} {(n+1)}\binom {n+1}h
(1+r)^{n-h+1}(1-r)^h\log\la_h.
\end{equation}
For $r=1$ this sum reduces to $\log\la_0$,
$\la_0$ being given by (\ref{e14}). A straightforward application of
Stirling's formula then leads to (\ref{a5}).

{}From now on let $0\le r<1$. We recall that $\la_h$ is given by
(\ref{e14}). Consequently, 
we expand the logarithm in (\ref{a6}) according to the addition rule, 
and split the sum (\ref{a6}) into the corresponding parts. The
individual parts can be summed by means of the binomial theorem,
except for the parts which involve $\log\Ga(1+h-u)$. (To be precise,
they have to be split appropriately before the binomial
theorem can be applied. Computer algebra systems like {\sl Maple} or
{\sl Mathematica} do this automatically.) 

In order to handle the terms which contain $\log\Ga(1+h-u)$, we use
Stirling's formula
\begin{equation} \label{eq:Stirling}
\log\Ga(z)=\(z-\frac {1} {2}\)\log(z)-z+\frac {1} {2}\log2
+\frac {1} {2}\log\pi+O\(\frac {1} {z}\).
\end{equation}
Again, after splitting, all the resulting sums can be evaluated by means of the
binomial theorem, except for
\begin{equation}\label{a7}
\frac {1} {2^{n+1}r}\sum _{h=0} ^{n+1}\frac {(n-2h+1)}
{(n+1)} \binom {n+1}h
(1+r)^{n+1-h}(1-r)^h (1/2-u+h)\log(1+h-u).
\end{equation}
The asymptotics of this sum can now easily (if though tediously) be
determined by making use of a Taylor expansion of $\log(1+h-u)$ about
$n(1-r)/2$ (i.e., at $1+h-u=n(1-r)/2$) with sufficiently many terms.

If everything is put together, the result is (\ref{a3}).
\quad \quad \qed
\medskip

\subsection{Asymptotics of the von Neumann
entropies of the Bayesian density
matrices $\ze_n(u)$}\label{s2.5}
The main result of this section describes the asymptotics of the
von Neumann
entropy (\ref{eq:1}) of $\ze_n(u)$. In view of the explicit description of
the eigenvalues of $\ze_n(u)$ and their multiplicities in
Theorem~\ref{t2}, this entropy equals 
\begin{equation} \label{eq:vonN}
-\sum _{h=0} ^{\fl{n/2}}\frac {(n-2h+1)^2} {(n+1)}\binom {n+1}h
\la_h\log\la_h,
\end{equation}
with $\la_h$ being given by (\ref{e14}).
\begin{theorem}\label{t11}
The asymptotics of the von Neumann entropy $S(\ze_n(u))$ of $\ze_n(u)$
is given by
\begin{multline}\label{B2}
 n \left( {\frac{-7 + 5 u} 
       {2 \left( 2 - u \right)  \left( 1 - u \right) }} + 
     {\psi}(5 - 2 u) - {\psi}(1 - u) \right)+
 \frac {3} {2}\log n+ 
  \left( -{\frac 7 2} + 2 u \right)  \log 2 \\-
{\frac{14 - 20 u + 7 {u^2}} 
    {2 \left( 2 - u \right)  \left( 1 - u \right) }}  + 
  \log \big({\Gamma}(1 - u)\big) - \log \big({\Gamma}({5/ 2} -
u)\big)\\
+(2-2u)(\psi(5-2u)-\psi(1-u))+O\(\frac {1} {n^{1-u}}\),
\end{multline}
where $\psi(x)$ is the digamma function,
$$\psi(x)=\frac {\frac {d} {dx}\Gamma(x)} {\Gamma(x)}.$$

\end{theorem}

{\sc Sketch of Proof}. 
We have to estimate the expression (\ref{eq:vonN}) for large $n$. We
proceed as in the proof of Theorem~\ref{t9}. First we use the
property $\la_{n+1-h}=\la_h$ to rewrite the sum (\ref{eq:vonN}) as
\begin{equation}\label{B11}
-\frac {1} {2}\sum _{h=0} ^{n+1}\frac {(n-2h+1)^2} {(n+1)}\binom {n+1}h
\la_h \log\la_h.
\end{equation}
Next, while recalling that $\la_h$ is given by (\ref{e14}), we 
expand the logarithm in (\ref{B11}) according to the addition rule, 
and split the sum (\ref{B11}) into the corresponding parts.
Here, the individual parts can be summed by means of Gau\ss' $_2F_1$
summation (\ref{e12}),
except for the parts which involve $\log\Ga(1+h-u)$. (Again, to be precise,
they have to be split appropriately before the Gau\ss{}
summation can be applied, which is done automatically by computer algebra 
systems like {\sl Maple} or {\sl Mathematica}.)

To handle the terms which contain $\log\Ga(1+h-u)$, we invoke again
Stirling's formula (\ref{eq:Stirling}).
After splitting, all the resulting sums can be evaluated by means of 
Gau\ss' $_2F_1$ summation (\ref{e12}), except for
\begin{equation}\label{B15} 
\sum _{h=0} ^{n+1}{\frac {(n-2h+1)^2} {(n+1)}\binom
{n+1}h}\la_h(1/2+h-u)\log(1+h-u).
\end{equation}
Now, to get an asymptotic estimate for this sum, as $n$ tends to
infinity, is not as obvious as it was for (\ref{a7}). The essential
``trick'' needed was kindly indicated to us by Peter Grabner: an asymptotic
estimate (in fact, an exact result) 
for (\ref{B15}) with $\log(1+h-u)$ replaced by $\psi(1+h-u)$
can be obtained without difficulty (but with some amount of
tedious calculation) by starting with the sum
\begin{multline}\label{B6}
\sum _{h=0} ^{n+1}\frac {(n-2h+1)^2} {(n+1)}\binom {n+1}h\\
\cdot
\frac {1} {2^n}\frac {\Ga(5/2-u)\,\Ga(2+n-h-u)\,\Ga(1+\al+h-u)}
{\Ga(5/2+n/2-u)\,\Ga(2+n/2-u)\,\Ga(1-u)}(h-u+1/2),
\end{multline}
evaluating it by applying Gau\ss' $_2F_1$ summation (\ref{e12}),
differentiating both sides of the resulting equation with respect to
$\al$, and by finally setting $\al=0$. Finally 
one relates the result to (\ref{B15}) by using the
asymptotic expansion
$\psi(z)=\log(z)-\frac {1} {2z}+O\(\frac {1}
{z^2}\)$.

If everything is put together, the right-hand side of (\ref{B2}) is
obtained.\quad \quad \qed
\medskip

\vskip10pt
\vbox{
\centerline{
\hbox{\psfig{file=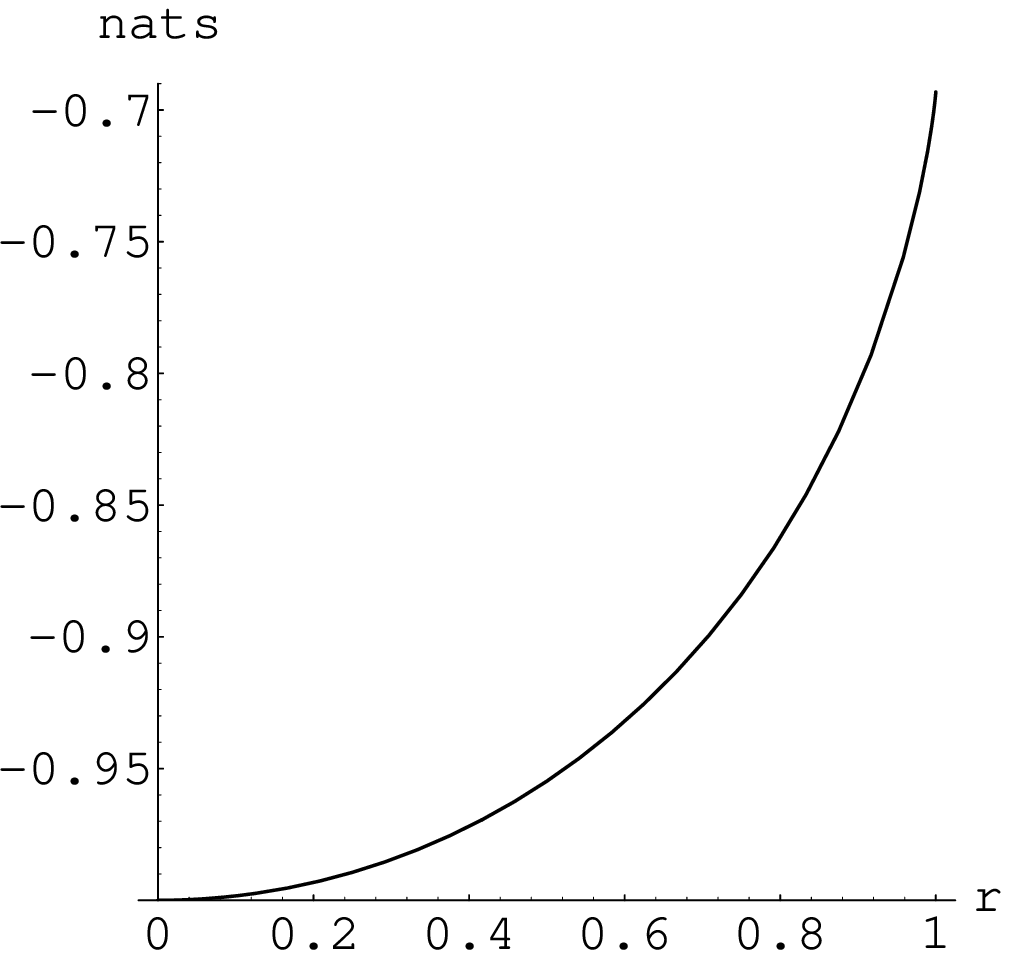,height=7cm}}
}
\centerline{\small Nonclassical/quantum term 
($\frac1  {2r} \big((1-r)\log(1-r)-(1+r)\log(1+r)\big)$)}
\centerline{\small in the quantum asymptotic redundancy (\ref{a3})}
\vskip5pt
\centerline{\small Figure 2}
}
\vskip10pt

\section{Comparison of our asymptotic redundancies
for the one-parameter family $q_{u}$ with those of Clarke
and Barron}\label{s3}
Let us, first, compare the formula (\ref{eq:4}) for the asymptotic redundancy
of Clarke and Barron to that derived here (\ref{a3}) for the two-level
quantum systems, in terms of the one-parameter family of probability
densities $q_{u}$, $ -\infty < u < 1$, given in (\ref{eq:10}).
Since the unit ball or Bloch sphere of such systems is three-dimensional
in nature, we are led to set the dimension $d$ of the parameter space in
(\ref{eq:4}) to 3. 
The quantum Fisher information matrix $I(\theta)$ 
for that case was taken to be (\ref{eq:8}), while the role
of the probability function $w(\theta)$ is played by $q_{u}$.
Under these substitutions, it was seen in the Introduction that 
formula (\ref{eq:4}) reduces to (\ref{eq:12}).
Then, we see that for 
$0 \le r < 1$, formulas (\ref{a3}) and
(\ref{eq:12})
coincide except for the presence of the
monotonically 
increasing (nonclassical/quantum) term 
$$-\frac {1} {2}\log(1-r^2)+\frac {1} {2r}\log\(\frac {1-r} {1+r}\)=
\frac1  {2r} \big((1-r)\log(1-r)-(1+r)\log(1+r)\big)$$
(see Figure~2 
for a plot of this term --- $\log 2 \approx .693147$ ``nats'' of
information equalling one ``bit'')
in (\ref{a3}). (This term would
have to be replaced by $-1$ --- that is,
its limit for $r \rightarrow 0$ --- to give (\ref{eq:12}).)
In particular, the order of magnitude, ${\frac3  2} \log n$,
is precisely the same in both formulas.
For the particular case $r=0$, the asymptotic 
formula (\ref{a3}) (see (\ref{a4}))
precisely coincides with (\ref{eq:12}).

In the case $r=1$, however, i.e., when we consider the
boundary of the parameter space (represented by the unit sphere), 
the situation is slightly tricky.
Due to the fact that the formula of Clarke and Barron 
holds only for interior points of the parameter space, we cannot expect
that, in general, our formula will resemble that of Clarke and Barron.
However, if the probability density, $q_{u}$, is concentrated on
the boundary of the sphere, then we may disregard the interior of the
sphere, and consider the boundary of the sphere as the {\em true}
parameter space. This parameter space is {\em two-dimensional\/} and
consists of interior points throughout.
Indeed, the probability density $q_{u}$ is concentrated on the
boundary of the sphere if we choose $u=1$ since, as
we remarked in the Introduction, in the limit $u\to1$, the
distribution determined by $q_{u}$ tends to the uniform distribution
over the boundary of the sphere. Let us, again, (naively) attempt to apply
Clarke and Barron's formula (\ref{eq:4}) to that case. We parameterize
the boundary of the sphere by polar coordinates 
$(\vartheta,\phi)$,
\begin{align}\notag x&=\sin\vartheta \cos\ph\\
y&=\sin\vartheta \sin\ph\notag\\
z&=\cos\vartheta,\notag\\
0\le \ph\le {}&2\pi,\ 0\le\vartheta\le\pi.\notag
\end{align}
The probability density induced by $q_{u}$ in the limit $u\to1$ then
is $\sin\vartheta/4\pi$, the density of the uniform distribution.
Using
\cite[eq.~(8)]{fuj} 
(see footnote~2), the quantum (symmetric logarithmic derivative)
Fisher information matrix turns out to be
\begin{equation}
\begin{pmatrix}1&0\\
0&\sin^2\vartheta\end{pmatrix}\quad,
\end{equation}
its determinant equalling, therefore, $\sin^2\vartheta$. So, 
setting $d=2$ and substituting
$\sin\vartheta/4\pi$ for $w(\th)$ and $\sin^2\vartheta$ for $I(\th)$ in
(\ref{eq:4})
gives $\log n+\log 2-1$. On the other hand, our formula (\ref{a5}),
for $u=1$, gives $\log n$. So, again, the terms differ only by a
constant. In particular, the order of magnitude is again the same.

Let us now focus our attention on the asymptotic minimax
redundancy (\ref{eq:3}) of Clarke and Barron. If in (\ref{eq:3}) we again set
$d$ to 3, we obtain (\ref{eq:11}), 
which, numerically, is $\frac {3} {2}\log n-1.96736+o(1)$.
Clarke and Barron prove that this minimax expression is only attained by the
(classical) Jeffreys' prior.
In order to derive its quantum counterpart
--- at least, a restricted (to the family $q_{u}$) version --- we
 have to determine the behavior of
\begin{equation}\label{d5}
\min_{-\infty<u<1}\max_{0\le r\le 1}S(\overset n\otimes \rho, \ze_n(u))
\end{equation}
for $n \rightarrow \infty$.
By Theorem~\ref{t9} we know that for large $n$ the relative entropy 
$S(\overset n\otimes \rho,\ze_n(u))$ equals
\begin{multline}\label{d6a} 
\frac {3} {2}\log n  -\frac {1} {2}- \frac {3} {2} \log 2  
-(1-u)\log(1-r^2)+\frac {1} {2r}\log\(\frac {1-r} {1+r}\)\\
+ \log\Ga(1 - u)  -\log\Ga(5/2 - u),
\end{multline}
up to an error of the order $O({1} /{n})$, which is uniform in $u$
and $r$ as long as $0\le r\le 1-\ep$ for any fixed $\ep>0$. Let us
for the moment ignore the error term. Then what we have to do is to
determine the minimax of the expression (\ref{d6a}), that is
\begin{equation}\label{d6b} 
\frac {3} {2}\log n  -\frac {1} {2}- \frac {3} {2} \log 2  +
\min_{-\infty<u<1}\max_{0\le r\le 1}f(r,u),
\end{equation}
where
\begin{equation} \label{d6c}
f(r,u)=-(1-u)\log(1-r^2)+\frac {1} {2r}\log\(\frac {1-r} {1+r}\)
+ \log\Ga(1 - u)  -\log\Ga(5/2 - u).
\end{equation}
This is an easy task. First of all, if $u<.5$ then the function
$f(r,u)$ is unbounded at $r=1$. Hence, to determine the minimax,
we can ignore that range of $u$. If $u=.5$, then $f(r,u)$ is maximal
at $r=1$, at which it attains the value 
$-\log 2+\frac {1}
{2}\log\pi\approx -0.120782$.
On the other hand, if $u>.5$ then
$f(r,u)$ attains a maximum in the interior of the interval $0<r<1$.
To determine this maximum,
we differentiate $f(r,u)$ with respect to $r$, to obtain
$$\frac{2 {r^2}-1} {r\left( 1 - r^2 \right) } - 
  \frac{2 r u} {1 - {r^2}} - 
  \frac{1} {2 {r^2}}\log \(\frac{1 - r} {1 + r}\).
$$
Equating this to 0 gives
\begin{equation} \label{d6d}
u=1 - \frac1 {2 {r^2}} - \frac{\left( 1 - {r^2} \right) 
      } {4 {r^3}}\log \(\frac{1 - r} {1 + r}\).
\end{equation}
Now we have to express $r$ in terms of $u$, $r=r(u)$ say,
substitute in $f(r,u)$, and determine
$\min_{-\infty<u<1}f(r(u),u)$. However, equivalently, we can express
$u$ in terms of $r$, $u=u(r)$ say (as was previously done in (\ref{d6d})),
substitute in $f(r,u)$, and determine 
$\min_{0\le r\le 1}f(r,u(r))$. In order to do so, we
differentiate $f(r,u(r))$ with respect to $r$, equate the result to
0, and solve for $r$. Numerically, the result is 
$r\approx .961574$. 
Substituting this back into (\ref{d6d}), we obtain 
$u\approx .542593$. The
value of $f(r,u)$ at these values of $r$ and $u$ is $-0.184320$.
This is smaller than that previously found  ($-0.120782$) for $u=.5$, so
that particular value of $u$ is not of concern for the minimax, as well.
 
In the beginning, we did ignore the error term. In fact, as is not
very difficult to see, since the error term is uniform in $u$
and $r$ as long as $0\le r\le 1-\ep$ for any fixed $\ep>0$, it is
legitimate to ignore the error term. To be precise, the asymptotic
minimax is the result above, subject to an error of $o(1)$, that is,
the value of (\ref{d6a}) for 
$r\approx .961574$ and 
$u\approx .542593$. This is
$\frac {3} {2}\log n-1.72404+o(1)$.
For $u=.5$, on the other hand, asymptotically, 
the maximum of the redundancy 
(\ref{e32}) (which, by the considerations above, 
is (\ref{d6a}) for $r=1$) equals ${\frac 3  2} \log n 
-\frac {1} {2}-\frac {5} {2}\log 2+\frac {1} {2}\log\pi+o(1)
\approx \frac {3} {2}\log n
-1.66050 +o(1)$.
We must, therefore, conclude that --- in contrast to the
classical case \cite{cl1,cl2} --- our trial candidate ($q_{0.5}$) for
the quantum counterpart of Jeffreys' prior does not exactly achieve the
minimax redundancy, although the prior $q_{0.542593}$ is remarkably
close to $q_{0.5}$, the hypothesized ``quantum Jeffreys' prior'' from
\cite{sl3,sl4}.

\smallskip
We now concern ourselves with the asymptotic {\it maximin} redundancy.
Clarke and Barron \cite{cl1,cl2} prove that the maximin redundancy is
attained asymptotically, again, by the Jeffreys' prior.
To derive the quantum counterpart of the maximin redundancy within
our analytical framework, we would have to calculate
\begin{equation}\label{d7}
\max _w\min_{Q_n}\int_{x^2+y^2+z^2\le 1}S(\overset
n\otimes\rho,Q_n)\,w(x,y,z)\,dx\,dy\,dz,
\end{equation}
where $Q_{n}$ varies over the $(2^{2 n}-1)$-dimensional convex set
of $2^n \times 2^n$ density matrices and $w$ varies over all
probability 
densities over the unit ball.
As we already mentioned in the Introduction,
in the classical case, due to a result of Aitchison 
\cite[pp.~549/550]{ait},
the minimum is achieved by setting
$Q_{n}$ to be the Bayes estimator,
i.e., the average of all possible 
probability densities in the family
that is considered with respect to the given
probability distribution.
In the quantum domain the same 
assertion is true. For the sake of completeness, we include the proof
in the Appendix. We can, thus, take the quantum
analog of the Bayes estimator 
to be the Bayesian density matrix $\zeta_{n}(u)$. 
That is, we set $Q_n=\ze_n(u)$ in (\ref{d7}).
Let us, for the moment, restrict the possible $w$'s over which
the maximum is to be taken to the family
$q_{u}$, $-\infty < u < 1$. Thus, we consider
\begin{equation} \label{d7a}
\max _u\int_{x^2+y^2+z^2\le 1} S(\overset
n\otimes\rho,\ze_n(u))\,q_{u}(x,y,z)\,dx\,dy\,dz. 
\end{equation}
By the definition (\ref{eq:5}) of relative entropy, we have
\begin{multline}\notag
S(\overset
n\otimes\rho,\ze_n(u))=\Tr\(\overset n\otimes\rho\log \overset
n\otimes\rho\) -\Tr\(\overset n\otimes\rho\log \ze_n(u)\)\\
=n{\frac{(1-r)}  2}{ \log {\frac{(1-r)}  2}} +n{\frac {(1+r) }
2}{ \log {\frac{(1+r)}   2}}-\Tr\(\overset n\otimes\rho\log
\ze_n(u)\),
\end{multline}
the second line being due to (\ref{eq:7}). Therefore, we get
\begin{multline} \label{d7b}
\int_{x^2+y^2+z^2\le 1} S(\overset
n\otimes\rho,\ze_n(u))\,q_{u}(x,y,z)\,dx\,dy\,dz\\
=\(n\int_0^1\int_0^{\pi}\int_0^{2\pi}\(
{\frac{(1-r)}  2}{ \log {\frac{(1-r)}  2}} +{\frac {(1+r) }
2}{ \log {\frac{(1+r)}   2}}\)\)r^2q_{u}\,d\ph\,d\vartheta\,dr\\
-\Tr\big(\ze_n(u)\log\ze_n(u)\big)\\
=- n \left( {\frac{-7 + 5 u} 
       {2 \left( 2 - u \right)  \left( 1 - u \right) }} + 
     {\psi}(5 - 2 u) - {\psi}(1 - u) \right)
+S(\ze_n(u)).
\end{multline}
{}From Theorem~\ref{t11}, we know the asymptotics of the von Neumann
entropy $S(\ze_n(u))$. Hence, we find that the
expression (\ref{d7b}) is asymptotically
equal to
\begin{multline} \label{d8}
 \frac {3} {2}\log n+ 
  \left( -{\frac 7 2} + 2 u \right)  \log 2 \\-
{\frac{14 - 20 u + 7 {u^2}} 
    {2 \left( 2 - u \right)  \left( 1 - u \right) }}  + 
  \log \big({\Gamma}(1 - u)\big) - \log \big({\Gamma}({5/ 2} -
u)\big)\\
+(2-2u)(\psi(5-2u)-\psi(1-u))+O\(\frac {1} {n^{1-u}}\).
\end{multline}
We have to, first, perform the maximization required in (\ref{d7a}),
and then determine the asymptotics of the result. 
Due to the form of the asymptotics in (\ref{d8}), we can, in fact,
derive the proper result by proceeding in the reverse order. That is,
we first determine the asymptotics of $\int S(\overset
n\otimes\rho,\ze_n(u))\,q_{u}\,dx\,dy\,dz$, which we did in (\ref{d8}),
and then we
maximize the $u$-dependent part in (\ref{d8}) with respect to $u$
(ignoring the error term).
(In Figure~3 we display this $u$-dependent part over the range
$[-0.2,1]$.)
Of course, we do the latter step by equating the first derivative of the
$u$-dependent part in (\ref{d8})
with respect to $u$ to zero and solving for $u$.
It turns out that this equation takes the appealingly simple form
\begin{equation} \label{maximin}
2(1-u)^3\big(\psi'(1-u)-\psi'(5/2-u)\big)=1.
\end{equation}
Numerically, we find  this equation to have the solution $u \approx .531267$,
at which the asymptotic maximin redundancy 
assumes the value ${\frac 3  2} \log n 
-1.77185+
O(1/n^{.468733})$. For $u=.5$, on the other hand, we have
for the asymptotic 
redundancy (\ref{d8}),
${\frac 3  2} \log n
-2-\frac {1} {2}\log 2+\frac {1}
{2}\log\pi+
O(1/\sqrt n)
\approx \frac {3} {2}\log n
-1.77421
+O(1/\sqrt n)$.
Again, we must, therefore, conclude that --- in contrast to the
classical case \cite{cl1,cl2} --- our trial candidate ($q_{0.5}$) for
the quantum counterpart of Jeffreys' prior can not
serve as a ``reference prior,'' in the sense introduced
by Bernardo \cite{berna,bern}. Moreover, --- again in contrast to the
classical situation \cite{hauss} --- 
we find that the minimax and the maximin are {\it not} identical (although
remarkably close). The two distinct priors yielding these values 
($q_{0.542593}$, respectively $q_{0.531267}$)
are themselves remarkably close, as well.

\vskip10pt
\vbox{
\centerline{
\psfig{file=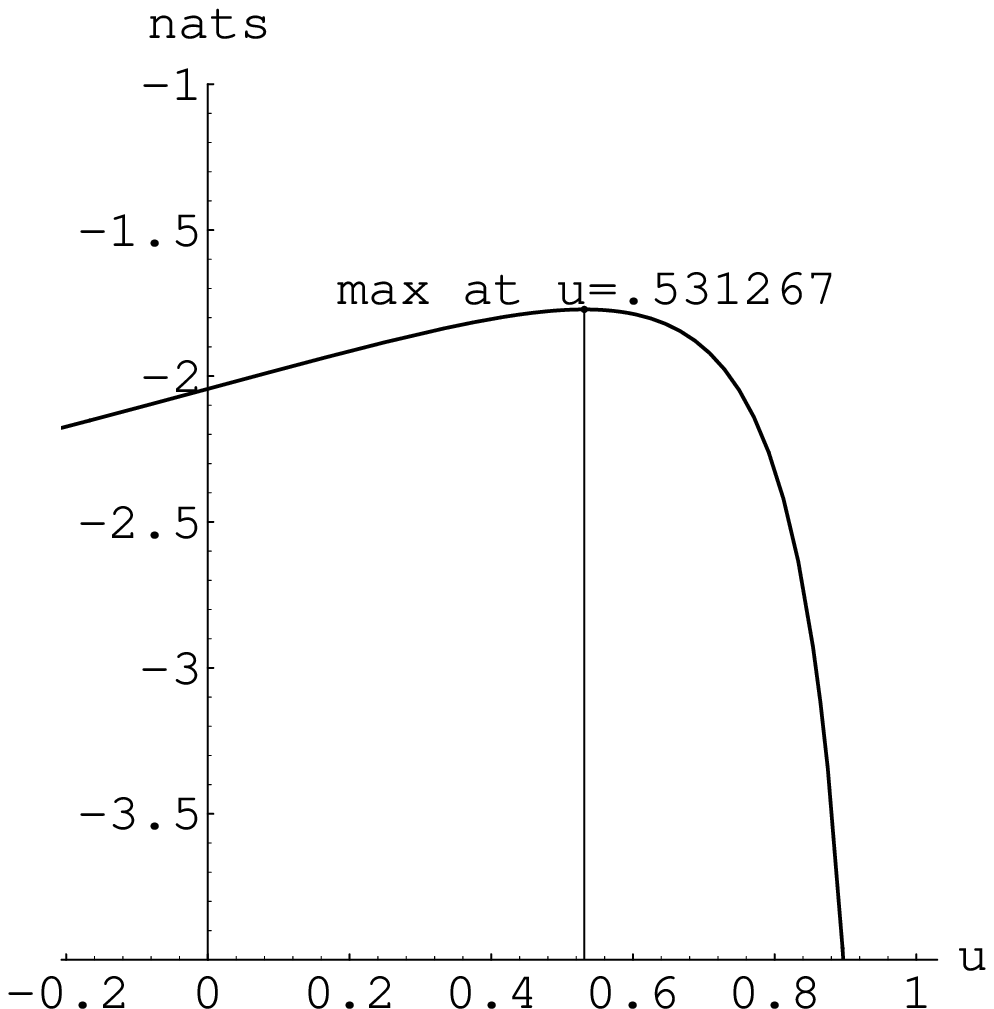,height=7cm}
}
\centerline{\small $u$-dependent part of the asymptotic Bayes
redundancy (\ref{d8})}
\vskip5pt
\centerline{\small Figure 3}
}
\vskip10pt

Since they are mixtures of product states, the matrices
$\zeta_{n}(u)$ are classically --- as opposed to EPR
(Einstein--Podolsky--Rosen) --- correlated
\cite{werner}. Therefore, $S(\zeta_{n}(u))$ must not be less than
the sum of the von Neumann entropies 
of any set of reduced
density matrices obtained from it, through computation of partial traces.
For positive integers, $n_{1} + n_{2}+ \cdots = n$, the corresponding
reduced density matrices are simply $\zeta_{n_{1}(u)}, \zeta_{n_{2}(u)},
 \dots$,
due to the mixing
\cite[Exercise~7.10]{belt}.
 Using these reduced density matrices, one can compute
{\it conditional} density matrices and quantum entropies \cite{cerf}.
Clarke and Barron \cite[p.~40]{cl1} have an alternative expression for
the redundancy in terms of conditional entropies, and it would be of
interest to ascertain whether a quantum analogue of this expression exists.

Let us note that the theorem of Clarke and Barron utilized the
uniform convergence property of the asymptotic expansion of the
relative entropy (Kullback--Leibler divergence). Condition 2 in their paper 
\cite{cl1} is, therefore,
crucial. It assumes --- as is typically the
 case classically --- that the matrix of
  second derivatives, $J(\theta)$, of the
relative entropy is identical to the Fisher information matrix $I(\theta)$.
In the quantum domain, however, in general, $J(\theta) \geq I(\theta)$,
where $J(\theta)$ is the matrix of second derivatives of the quantum relative 
entropy (\ref{eq:5}) and $I(\theta)$ is the symmetric logarithmic
derivative Fisher information matrix \cite{pe1,pe2}.
The equality holds only for special cases. For instance,
$J(\theta) > I(\theta)$ does hold if $r \neq 0$ for the situation
considered in this paper.
The volume element of the Kubo-Mori/Bogoliubov (monotone) metric \cite{pe1,pe2}
is given by $\sqrt {\det J(\theta)}$. This can be normalized
for the two-level quantum systems to be a member
($u=1/2$) of a one-parameter family of probability 
densities
\begin{equation} \label{eq:Kubo}
\frac{(1-u) \,\Gamma(5/2-u) \,r\, \log{\big((1+r)/(1-r)\big)}\, \sin \vartheta}
{\pi^{3/2}\, (3 - 2 u)\, \Gamma(1-u)\, (1-r^2)^u},\quad -\infty<u<1,
\end{equation}
and similarly studied, it is presumed,
in the manner of the family $q_{u}$ (cf.~(\ref{eq:10}) and (\ref{e7})) analyzed 
here. These
two families can be seen to differ --- up to the normalization
factor --- by the replacement of $\log {\big((1+r)/(1-r)\big)}$ in (\ref{eq:Kubo})
by, simply, $r$.
(These two last expressions are, of course, equal for $r=0$.)
In general, the volume element of a monotone metric over the two-level
quantum systems is of the form \cite[eq.~3.17]{pe1}
\begin{equation} \label{monotone}
\frac{r^2 \sin{\vartheta}}{f\big((1-r)/(1+r)\big) (1-r^2)^{1/2} (1+r)},
\end{equation}
where $f:\mathbb R^+ \rightarrow \mathbb R^+$ is an operator monotone function
such that $f(1) = 1$ and $f(t) = tf(1/t)$.
For $f(t)=(1+t)/2$, one recovers the volume element
($\sqrt {\det {I(\theta)}}$) of the
metric of the symmetric
logarithmic derivative, and for $f(t) = 
(t-1)/ {\log t}$, that
($\sqrt {\det {J(\theta)}}$) of the
Kubo-Mori/Bogoliubov metric \cite{pe3,pe1,pe2}.
(It would appear, then, that the only member of the family $q_{u}$
proportional to a monotone metric is $q_{0.5}$, that is (\ref{eq:9}).
 The maximin result
we have obtained above corresponding to $u \approx .531267$ --- the solution
of (\ref{maximin}) --- would
 appear unlikely, then, to extend globally beyond the family.
Of course, a similar remark could be made in regard to to the minimax,
corresponding to $u \approx .542593$, as shown above.)
While $J(\theta)$ can be generated from the relative entropy (\ref{eq:5})
(which is a limiting case of the $\alpha$-entropies \cite{pe4}),
$I(\theta)$ is similarly obtained from \cite[eq.~3.16]{pe3}
\begin{equation} \label{jan}
 \Tr \rho_{1} (\log \rho_{1} - \log \rho_{2})^2 .
\end{equation}
It might prove of interest to repeat the general line of
analysis carried out in this paper, but with the use
of (\ref{jan}) rather than (\ref{eq:5}).
Also of importance might be an analysis in which the relative
entropy (\ref{eq:5}) is retained, but the family (\ref{eq:Kubo}) based
on the Kubo-Mori/Bogoliubov metric is used instead of $q_{u}$.
Let us also indicate that if one equates the asymptotic
redundancy formula of Clarke and Barron (\ref{eq:4})
(using $w(\theta) = q_{u}(x,y,z)$) to that derived here
(\ref{a3}),
 neglecting the residual terms, solves for
$\det(I(\theta))$, and takes the square root of the result,
one obtains a prior of the form (\ref{monotone})
 based on the
monotone function 
$f(t) =t^{t/(t-1)}/e$. (Let us note
that the reciprocal of the related ``Morozova-Chentsov'' function \cite{pe1},
$c(x,y) = 1/y f(x/y)$, in this case, is the {\it exponential} mean \cite{qi}
of $x$ and $y$, while for the minimal monotone metric, the reciprocal of
the Morozova-Chentsov function is the {\it arithmetic} mean. It is, therefore,
quite interesting from an information-theoretic point of view that these are,
in fact, 
the only 
two means which furnish additive quasiarithmetic average codeword
lengths \cite[p. 157]{aczel}. Also, it appears to be a quite important, 
challenging question --- bearing upon the relationship
between classical and quantum probability --- to determine
 whether or not a family of 
probability distributions over the Bloch sphere exists, which yields
as its volume
element for  the corresponding Fisher information matrix, a prior of the
form (\ref{monotone}) with the noted $f(t) = t^{t/(t-1)}/e$.)

As we said in the Introduction, ideally we would like to start with a
(suitably well-behaved) {\it arbitrary\/} probability density
on the unit ball, determine the relative entropy of 
$\overset n\otimes \rho$ with respect to the average of 
$\overset n\otimes \rho$ over the probability density,
then find its asymptotics, and finally, among all such probability
densities, find the one(s) for which the minimax and maximin are
attained. In this regard, we wish to mention that a suitable
combination of results and computations from Sec.~\ref{s2} with basic
facts from representation theory of $SU(2)$ 
(cf\@. \cite{vilklim,biedlouck} for more information on that topic) 
yields the following result.

\begin{theorem}\label{t15}
Let $w$ be a spherically symmetric probability density on the unit
ball, i.e., $w=w(x,y,z)$ depends only on $r=\sqrt{x^2+y^2+z^2}$.
Furthermore, let $\hat \ze_n(w)$ be the average 
$\int_{x^2+y^2+z^2\le 1} \big(\overset n  \otimes \rho \big)\,w \,dx\,dy\,dz$.
Then the eigenvalues of 
$\hat\ze_n(w)$ are
\begin{equation}\label{e50}
\la_{h}=\frac {\pi} {2^{n-1}(n-2h+1)}\int_{-1}^{1}
r(1+r)^{n-h+1}(1-r)^h w(\v{r})\,dr, 
\quad h=0,1,\dots,\fl{\frac {n} {2}},
\end{equation}
with respective multiplicities
\begin{equation}\label{e51}
\frac {(n-2h+1)^2} {(n+1)}\binom {n+1}h,
\end{equation}
and corresponding eigenspaces 
$\{v_{h,s}(P):h\le s\le n-h,\ P$ a ballot path from $(0,0)$
to $(n,n-2h)\}$, which were described in Sec.~\ref{s2.2}.

The relative entropy of $\overset n  \otimes \rho$ with respect to
$\hat\ze_n(w)$ is given by {\em(\ref{e32})}, with $\la_h$ as given in
{\em(\ref{e50})}.
\end{theorem}
We hope that this Theorem enables us to determine the
asymptotics of the relative entropy and, eventually, to 
find, at least within the family of spherically
symmetric (that is, unitarily-invariant)
 probability densities on the unit ball, the corresponding
minimax and maximin redundancies. Doing so, would resolve the outstanding
question of whether these two redundancies, in fact, coincide, as classical
results would suggest \cite{hauss}.

\section{Summary}\label{s4}
Clarke and Barron \cite{cl1,cl2} (cf\@. \cite{ri}) have derived several
forms of asymptotic redundancy for arbitrarily parameterized
families of probability distributions.
We have been motivated to undertake this study by the possibility
that their results may generalize, in some yet not fully understood
fashion, to the quantum domain of noncommutative probability.
(Thus, rather than probability 
densities, we have been concerned
here with density matrices.)
We have only, so far, been able to examine this possibility in a
somewhat restricted manner.
By this, we mean that we have limited our consideration to two-level
quantum systems (rather than $n$-level ones, $n \geq 2$), and for the case
$n=2$, we have studied (what has proven
to be) an analytically tractable one-parameter family of possible prior
probability 
densities, $q_{u}$, $-\infty <u<1$
(rather than the totality of arbitrary probability 
densities).
Consequently, our results can not be as definitive in
nature as
those of Clarke and Barron.
Nevertheless, the analyses presented here reveal
  that our trial candidate ($q_{0.5}$, that is (\ref{eq:9})) for the quantum 
counterpart of the Jeffreys' prior
 closely approximates those probability distributions
which we have, in fact, found to yield the minimax ($q_{0.542593}$) and 
maximin ($q_{0.531267}$) for our one-parameter family
($q_u$).

                                              Future research
might be devoted to expanding the family of probability distributions used
to generate the Bayesian density matrices for $n=2$, as well as
similarly studying the $n$-level quantum systems ($n>2$).
(In this
regard, we have examined the situation in which $n=2^m$, and the
only $n \times n$
density matrices considered are simply the tensor products of $m$ identical
$2 \times 2$ density matrices. Surprisingly,
for $m=2,3$, the associated trivariate
candidate  quantum Jeffreys' prior,
taken, as throughout this study,
 to be proportional to the volume elements of the metrics
of the symmetric logarithmic derivative (cf\@. \cite{sl4}),
have been found to be
   {\it improper} (nonnormalizable) over the
Bloch sphere. The minimality of such metrics is guaranteed, however,
only if ``the whole state space of a spin is parameterized'' \cite{pe1}.)
In all such cases, it will be of interest to evaluate the characteristics
of the relevant candidate quantum Jeffreys' prior {\it vis-\`a-vis} all other
members of the family
of probability distributions employed over the
$(n^2-1)$-dimensional
convex set of
$n \times n$ density matrices.

We have also conducted analyses parallel to those
reported above, but having,
{\it ab initio},
 set either $x$ or $y$ to zero in
the $2 \times 2$ density matrices (\ref{eq:6}). This,
then, places us in the realm
of real --- as opposed to complex (standard or conventional)
quantum mechanics. (Of course, setting {\it both} $x$ and $y$
to zero would return us to a strictly classical situation, in which
the results of Clarke and Barron \cite{cl1,cl2}, as applied to binomial
distributions, would be directly applicable.)
Though we have --- on the basis
of detailed computations --- developed strong conjectures as to the nature
of the 
associated results, we have not,
at this stage of our investigation, yet succeeded in formally
demonstrating their validity.

In conclusion, again in analogy to classical results,
we would like to raise the possibility that the quantum
asymptotic redundancies derived here might prove of
value in deriving formulas for the {\it stochastic
complexity} \cite{ri,ri2} 
(cf.~\cite{svozil}) --- the shortest
description length --- of a string of $n$
{\it quantum} bits. The competing possible models for the data string
might be taken to be the $2 \times 2$ density matrices ($\rho$)
corresponding to different values of $r$, or equivalently,
different values of the von Neumann entropy, $S(\rho)$.

\section*{Appendix: The quantum Bayes estimator achieves the minimum
average entropy} 
Let $P_\th$, $\th\in\Th$, be a family of density matrices, and let
$w(\th)$, $\th\in\Th$, be a probability density on $\Th$. 

\begin{theorem}
The minimum
$$\min_Q\int w(\th)S(P_\th,Q)\,d\th,$$
taken over all density matrices $Q$, is achieved by $M=\int w(\th)P_\th\,d\th$.
\end{theorem}
{\sc Proof}. We look at the difference
$$\int w(\th)S(P_\th,Q)\,d\th-\int w(\th)S(P_\th,M)\,d\th,$$
and show that it is nonnegative. Indeed,
\begin{align*}
\int w&(\th)S(P_\th,Q)\,d\th-\int w(\th)S(P_\th,M)\,d\th\\&=
\int w(\th)\Tr(P_\th\log P_\th-P_\th\log Q)\,d\th-
\int w(\th)\Tr(P_\th\log P_\th-P_\th\log M)\,d\th\\
&=\int w(\th)\Tr\big(P_\th(\log M -\log Q)\big)\,d\th\\
&=\Tr\Big(\Big(\int w(\th)P_\th\,d\th\Big)(\log M -\log Q)\Big)\\
&=Tr\big(M(\log M -\log Q)\big)\\
&=S(M,Q)\ge0,
\end{align*}
since relative entropies 
of density matrices are nonnegative \cite[bottom of p.~17]{oh}. 
\quad \quad \qed

\medskip

\section*{Acknowledgments}
Christian Krattenthaler did part of this research at the 
Mathematical Sciences Research Institute, Berkeley, during the
Combinatorics Program 1996/97.
Paul Slater would like to express appreciation
to the Institute for Theoretical Physics for
computational support. This research was undertaken, in part, to respond to
concerns (regarding the rationale for the
presumed quantum Jeffreys' prior)
 conveyed to him by Walter Kohn
and members of the informal seminar group he leads.
The co-authors are grateful to: Ira Gessel for
bringing them into initial contact {\it via} the Internet;
to Helmut Prodinger and Peter Grabner for their hints regarding 
the asymptotic computations; 
to A. R. Bishop and an anonymous referee of \cite{sl4}; and to the two
anonymous referees of this paper itself,
whose comments helped to considerably improve the presentation.

\end{document}